\documentclass[iop,apj]{emulateapj}
\usepackage{times,mathptmx}
\usepackage{amsmath,amssymb}
\begin{document}

\shorttitle{GALEX far-UV color selection of UV-bright high-redshift quasars}
\shortauthors{Worseck \& Prochaska}
\slugcomment{accepted for publication in ApJ}

\title{GALEX far-UV color selection of UV-bright high-redshift quasars}

\author{G\'abor Worseck and J.~Xavier Prochaska}
\affil{Department of Astronomy and Astrophysics, UCO/Lick Observatory,
University of California, 1156 High Street, Santa Cruz, CA 95064}
\email{gworseck@ucolick.org, xavier@ucolick.org}

\begin{abstract}
We study the small population of high-redshift ($z_\mathrm{em}>2.7$) quasars
detected by GALEX, whose far-UV emission is not extinguished by intervening
\ion{H}{1} Lyman limit systems. These quasars are of particular importance to
detect intergalactic \ion{He}{2} absorption along their sightlines. We correlate
almost all verified $z_\mathrm{em}>2.7$ quasars to the GALEX GR4 source catalog
covering $\sim 25000$\,deg$^2$, yielding 304 sources detected at S/N$>3$.
However, $\sim 50\%$ of these are only detected in the GALEX NUV band,
signaling the truncation of the FUV flux by low-redshift optically thick Lyman
limit systems. We exploit the GALEX UV color $m_\mathrm{FUV}-m_\mathrm{NUV}$ to
cull the most promising targets for follow-up studies, with blue (red) GALEX colors
indicating transparent (opaque) sightlines. Extensive Monte Carlo simulations
indicate a \ion{He}{2} detection rate of $\sim 60$\% for quasars with
$m_\mathrm{FUV}-m_\mathrm{NUV}\la 1$ at $z_\mathrm{em}\la 3.5$,
a $\sim$50\% increase over GALEX searches that do not include color information.
We regard 52 quasars detected at S/N$>3$ to be most
promising for HST follow-up, with an additional 114 quasars if we consider S/N$>2$
detections in the FUV. Combining the statistical properties of \ion{H}{1} absorbers
with the SDSS quasar luminosity function, we predict a large all-sky population of
$\sim 200$ quasars with $z_\mathrm{em}>2.7$ and $i\la 19$ that should be detectable
at the \ion{He}{2} edge at $m_\mathrm{304}<21$. However, SDSS provides just half of
the NUV-bright quasars that should have been detected by SDSS \& GALEX. With mock
quasar photometry we revise the SDSS quasar selection function, finding that SDSS
systematically misses quasars with blue $u-g\la 2$ colors at
$3\la z_\mathrm{em}\la 3.5$ due to overlap with the stellar locus in color space.
Our color-dependent SDSS selection function naturally explains the inhomogeneous
$u-g$ color distribution of SDSS DR7 quasars as a function of redshift and the
color difference between color-selected and radio-selected SDSS quasars.
Moreover, it yields excellent agreement between the observed and the predicted
number of GALEX UV-bright SDSS quasars. We confirm our previous claims that SDSS
preferentially selects $3\la z_\mathrm{em}\la 3.5$ quasars with intervening
\ion{H}{1} Lyman limit systems. Our results imply that broadband optical color
surveys for $3\la z_\mathrm{em}\la 3.5$ quasars have likely underestimated their
space density by selecting IGM sightlines with an excess of strong \ion{H}{1} absorbers.
\end{abstract}


\section{Introduction}

The intergalactic space is pervaded by a filamentary cosmic web of gas of almost
primordial composition, the so-called intergalactic medium (IGM), seen in
absorption against background sources \citep{rauch98,meiksin09}. The absence of
\ion{H}{1} Ly$\alpha$ absorption troughs in spectra of $z_\mathrm{em}<6$ quasars
signals that the hydrogen in the IGM is highly ionized \citep{gunn65}. Instead,
the plethora of narrow \ion{H}{1} Ly$\alpha$ absorption lines, known as the
Ly$\alpha$ forest, traces the tiny residual neutral hydrogen fraction of the
IGM as the largest reservoir of baryons in the universe. The ionizing radiation
of quasars and star-forming galaxies is filtered by the IGM, leading to the
buildup of the UV background radiation field that determines the ionization state
of the gas \citep{haardt96,fardal98,faucher09}. The UV background changes in
amplitude and spectral shape due to evolution in the source number density,
cosmological expansion and structure formation \citep[e.g.][]{dave99}. This is
particularly important for the ionization state of helium, the second most
abundant element in the IGM. Due to its 5.4 times higher recombination rate
and 4 times higher ionization threshold, the reionization epoch of helium
(\ion{He}{2}$\longrightarrow$\ion{He}{3}) is expected to be delayed with respect
to hydrogen.

The Ly$\alpha$ transition of intergalactic \ion{He}{2} at
$\lambda_\mathrm{rest}=303.78$\AA\ is observable in the far UV (FUV) from space
only at $z>2$ due to the Galactic Lyman limit. The determination of the
\ion{He}{2} reionization epoch via the \ion{He}{2} Gunn-Peterson test towards
high-redshift quasars has been a major goal in extragalactic UV astronomy since
the launch of the Hubble Space Telescope
\citep[HST, e.g.][]{miralda-escude90,miralda-escude93}. However, the accumulated
Lyman continuum (LyC) absorption of the \ion{H}{1} absorber population severely
attenuates the quasar flux in the FUV, rendering just a few percent of
$z_\mathrm{em}>3$ sightlines to be relatively transparent \citep{moller90}.
The combination of the rising LyC absorption and the declining quasar luminosity
function results in a sharply dropping number of observable UV-bright quasars at
$z_\mathrm{em}>3$ \citep{picard93,jakobsen98}.

Until very recently \ion{He}{2} Ly$\alpha$ absorption had been found only in a
handful of sightlines despite considerable effort, since the UV fluxes of most
targeted quasars had been unknown. HST observations of Q~0302$-$003 at
$z_\mathrm{em}=3.285$ \citep{jakobsen94,hogan97,heap00} and PKS~1935$-$692 at
$z_\mathrm{em}=3.18$ \citep{anderson99} revealed a high \ion{He}{2} effective
optical depth at $z\ga 3$ that is consistent with a Gunn-Peterson trough
($\tau_\mathrm{eff,He\,II}>3$). In contrast, the lines of sight towards
HS~1700$+$6416 at $z_\mathrm{em}=2.736$ \citep{davidsen96,fechner06},
HE~2347$-$4342 at $z_\mathrm{em}=2.885$ \citep{reimers97,kriss01,smette02,zheng04,shull04}
and HS~1157$+$3143 at $z_\mathrm{em}=2.989$ \citep{reimers05} show patchy
\ion{He}{2} absorption with voids ($\tau_\mathrm{eff,He\,II}<1$) and troughs
($\tau_\mathrm{eff,He\,II}>3$). At $z\la 2.7$ this patchy absorption evolves
into a \ion{He}{2} Ly$\alpha$ forest that has been resolved in high-resolution
spectra obtained with the Far Ultraviolet Spectroscopic Explorer
\citep[FUSE, ][]{kriss01,zheng04,shull04,fechner06}.

The strong evolution of the \ion{He}{2} absorption suggests a late reionization
epoch of helium at $z\sim 3$, when quasars have been sufficiently abundant to
supply the required hard photons. The patch-work of absorption and transmission
evokes a picture of overlapping \ion{He}{3} zones around quasars that lie close
to the sightline \citep{reimers97,heap00,smette02}. Indeed, the \ion{He}{3}
proximity zones of quasars have been detected both along the line of sight
\citep{hogan97,anderson99} and in transverse direction \citep{jakobsen03}. In
the past few years, great progress has been made in developing the theoretical
framework to interpret these observations. Both semi-analytic
\citep[e.g.][]{haardt96,fardal98,gleser05,furlanetto10} and numerical radiative
transfer simulations \citep{maselli05,tittley07,paschos07,mcquinn09a} indicate
that the \ion{He}{2} reionization process should be very inhomogeneous and
extended over $3\la z\la 4$, since rare luminous quasars dominate the
photoionizing budget of the overall quasar population. The few quasars
contributing to the UV radiation field at the \ion{He}{2} ionization edge at a
given point likely give rise to fluctuations in the FUV background that can be
tracked by the co-spatial absorption of \ion{He}{2} and \ion{H}{1}
\citep{bolton06,worseck06,worseck07,furlanetto09}. The UV background hardens
as \ion{He}{2} reionization proceeds \citep{heap00,zheng04}, but $\ga 10$~Mpc
fluctuations are expected to persist even after its end \citep{fechner07}.

Other, more indirect observations might suggest that \ion{He}{2} reionization
is ending at $z\sim 3$. The IGM is reheated as the individual \ion{He}{3}
bubbles around quasars overlap, however the amplitude of this temperature jump
is highly uncertain \citep{bolton09,bolton09b,mcquinn09a}. Observationally,
several studies indicated a jump in the IGM temperature at $z\sim 3$
\citep{ricotti00,schaye00,theuns02b}, whereas others are consistent with an
almost constant IGM temperature at $2\la z\la 4$ \citep{mcdonald01b,lidz10}.
Moreover, photoionization models of metal line systems indicate a significant
hardening of the UV background at $z\la 3$ \citep{agafonova05,agafonova07}.
However, these observations are restricted to rare metal line systems showing
various ions with a simple velocity structure.

At present, the five \ion{He}{2} absorption sightlines studied at scientifically
useful spectral resolution provide the best observational constraints on
\ion{He}{2} reionization. However, just one or two sightlines probe the same
redshift range, and given the large predicted variance in the \ion{He}{2}
absorption, this small sample clearly limits our current understanding of
\ion{He}{2} reionization\footnote{Ironically, the $z\sim 6$ epoch has
substantially better statistics.}. The Sloan Digital Sky Survey (SDSS) has
dramatically increased the number of high-redshift quasars to search for the
presence of flux at \ion{He}{2} Ly$\alpha$, yielding three $z_\mathrm{em}>3.5$
quasars with detected \ion{He}{2} Gunn-Peterson troughs
\citep{zheng04b,zheng05,zheng08}. More importantly, the almost completed first
UV all-sky survey with the Galaxy Evolution Explorer (GALEX) enables the
pre-selection of UV-bright quasars for follow-up UV spectroscopy, leading to
the recent discovery of 22 new clear sightlines towards SDSS quasars at
$3.1<z_\mathrm{em}<3.9$ \citep{syphers09b,syphers09a}. The available GALEX
photometry dramatically increases the survey efficiency by almost an order of
magnitude to $\simeq 42$\% in the \citeauthor{syphers09b} survey.

The recently installed Cosmic Origins Spectrograph (COS) on HST offers
unprecedented sensitivity to study \ion{He}{2} reionization via \ion{He}{2}
Ly$\alpha$ absorption spectra. With its confirmed throughput at
$\lambda>1105$\AA\ \citep{mccandliss10} HST/COS is now able to probe \ion{He}{2}
Ly$\alpha$ at $z>2.64$, thereby covering the full redshift range of interest for
\ion{He}{2} reionization. Very recently, \citet{shull10} presented a high-quality
COS spectrum of HE~2347$-$4342, dramatically improving on earlier FUSE data.
In the near future, COS will be employed to both obtain follow-up spectroscopy
of the recently confirmed \ion{He}{2} sightlines, and to discover new ones.
In this paper we introduce the quasar UV color measured by GALEX as a powerful
discriminator to select the most promising sightlines for follow-up spectroscopy.
Moreover, we significantly improve on earlier predictions on the number of UV-bright
quasars \citep{picard93,jakobsen98}, based on observational advances to characterize
both the quasar luminosity function and the optically thick IGM absorber
distribution. The structure of the paper is as follows: In \S\ref{galexsample}
we will present our sample of verified high-redshift quasars detected by GALEX.
Section \ref{mcspectra} describes our Monte Carlo routine to compute \ion{H}{1}
absorption spectra and to perform mock GALEX and SDSS photometry. In
\S\ref{results} we determine the expected number of UV-bright $z_\mathrm{em}>2.7$
quasars and establish GALEX UV color selection criteria to select quasars with
probable \ion{He}{2}-transparent sightlines. We compare the observed and
predicted number counts of UV-bright SDSS quasars in \S\ref{sdssgalex} before
concluding in \S\ref{conclusions}.

\section{Our sample of $z\ge 2.7$ quasars detected by GALEX}
\label{galexsample}

\subsection{The initial quasar sample}

We compiled a list of practically all known quasars at $z_\mathrm{em}\ge 2.7$
from four quasar samples. We started with the SDSS DR5 quasar catalog
\citep{schneider07} and added all other spectroscopic SDSS targets from DR6
\citep{adelman08} and DR7 \citep{abazajian09} identified as
$z_\mathrm{em}\ge 2.7$ quasars by the SDSS spectro1d pipeline. We supplemented
this SDSS quasar list by all $z_\mathrm{em}\ge 2.7$ sources from the
\citet{veron06} catalog not discovered or verified by SDSS. This merged quasar
catalog is inhomogeneous due to several reasons: (i) the SDSS DR5 quasar catalog
represents a non-statistical sample due to changes in the quasar selection
criteria in the course of the SDSS \citep{richards06,schneider07}, (ii) the
inclusion of SDSS quasars discovered by serendipity \citep{stoughton02}, (iii)
the redshifts of most SDSS DR6/7 sources have not been verified by eye, and
(iv) the \citet{veron06} catalog is inherently inhomogeneous as it is a
collection of quasars discovered by various surveys with sometimes unknown
selection criteria.

The merged list of quasars contained 12373 unique entries. However, among them
there are SDSS DR6/7 sources misidentified as high-$z$ quasars by the SDSS source
identification algorithm either due to misclassification or a wrong redshift assignment.
We refrained from the tedious visual classification of all spectro1d DR6/7 quasars
(see \citealt{schneider10} for the DR7 quasar catalog compiled after our analysis was finished),
and limited our visual verification to the subset of SDSS DR6/7 sources actually
detected by GALEX (see below). Moreover, we caution
that the \citet{veron06} catalog contains a fair number of quasar candidates
with estimated redshifts from slitless spectroscopic surveys. Many of these
redshifts will be grossly overestimated as most slitless spectroscopic surveys
assign the highest plausible redshifts if just a single emission line is
present. Consequently, we removed all misidentified SDSS sources and all quasar
candidates without unambiguous redshifts from follow-up spectroscopy, but only
after cross-correlating the initial quasar sample to the GALEX GR4 source catalog.

\subsection{Cross-correlation with GALEX GR4}

The GALEX satellite currently performs the first large-scale UV imaging survey
\citep{martin05,morrissey07}. Most images are taken simultaneously in two
broad bands, the near UV (NUV, $\sim$1770--2830\AA) and the far UV
(FUV, $\sim$1350--1780\AA) at a resolution of $\sim 5\arcsec$ full width
at half maximum (FWHM). Three nested GALEX imaging surveys have been defined:
the All-Sky Survey (AIS) covering essentially the whole extragalactic sky
($\sim$26000\ deg$^2$) to $m_\mathrm{AB}\sim 21$, the Medium Imaging Survey (MIS)
reaching $m_\mathrm{AB}\sim 23$ on 1000\ deg$^2$, and the Deep Imaging Survey (DIS)
extending to $m_\mathrm{AB}\sim 25$ on 80\ deg$^2$. These main surveys are
complemented by guest investigator programs. The GALEX Data Release 4 (GR4)
covers $\sim$25000\ deg$^2$, 96\% of the anticipated AIS survey area. The
officially distributed GR4 data has been homogeneously reduced and analyzed by
a dedicated software pipeline. A previous version of this pipeline used for the
earlier GR3 data release is described in detail by \citet{morrissey07}.

We cross-correlated our initial quasar list to the available GALEX GR4 source
catalogs using a maximum match radius of 4.8\arcsec\ around the optical quasar
position. The match radius approximately corresponds to the typical GALEX FWHM
and was chosen to account for the degrading astrometric accuracy of GALEX towards
the detection limit where we expect most of the rare UV-transparent quasars
(see \S\ref{sect_verif} below). In comparison, the positional errors of the
quasars are negligible, $0\farcs1$ for SDSS \citep{pier03} and $\la 1\arcsec$
for the \citeauthor{veron06} catalog quasars.

\subsection{Source verification and catalog completeness}
\label{sect_verif}

Substantial screening of the cross-matches was required to create our final list
of real $z_\mathrm{em}\ge 2.7$ quasars detected in GALEX GR4. We visually
confirmed the redshift of every detected SDSS source and searched the references
of the \citeauthor{veron06} catalog quasars for unambiguous redshift
determinations and plotted spectra. A large fraction of the GALEX-detected
\citeauthor{veron06} quasars had unconfirmed slitless spectroscopic redshifts,
in line with our assertion that most of them are in fact low-redshift
interlopers. Consequently we removed these unconfirmed candidates. In addition,
we flagged obvious broad-absorption-line (BAL) quasars which are rarely usable
for IGM studies due to the difficulty in disentangling the IGM absorption along
their sightlines from the high-velocity quasar outflows. This flagging was
somewhat restrictive, as it was based on the visual appearance of the spectrum
(if available), and quasars with confined low-velocity narrow BAL systems were
kept in the sample. Finally, we inspected the SDSS images of all GALEX-detected
quasars in the SDSS DR7 footprint, and flagged cases of potential source
confusion with blue optical neighbors at $\la 5\arcsec$ separation caused by the
broad GALEX point spread function (PSF). Specifically, a quasar was flagged if the
spectral energy distribution of the neighbor (as estimated from the SDSS photometry)
was likely to extend to the UV (e.g.\ significant $u$ band flux). In total,
$\simeq$20\% of the SDSS quasars were flagged. Lacking deep multi-band
photometry, we could not inspect the \citeauthor{veron06} quasars outside of the
SDSS footprint with the same scrutiny. For quasars imaged in multiple GALEX
exposures we kept only the most significant detection, usually in the deepest
exposure unless affected by obvious image artifacts. For every source formally
detected in only one GALEX band we obtained a $1\sigma$ upper limit on the flux
in the other. In total, we were left with 803 verified $z_\mathrm{em}>2.7$
quasars with likely GALEX GR4 counterparts. Almost all of them (782) have been imaged
in both GALEX filters, allowing for constraints on the UV color (\S\ref{obscolors}).

Due to the strong Lyman continuum absorption by the intervening IGM most of
these high-redshift quasars are faint in the UV even if they are optically bright
(see \S\ref{results_counts} below). Most of these rare high-redshift quasars with
appreciable UV flux will be detected at low signal-to-noise (S/N) close to the
limits of the defined GALEX imaging surveys.
Incompleteness arises in the source catalog at low S/N, resulting in false
negatives (nondetections in one or both bands) and false positives
(no UV flux at all). The low-S/N UV fluxes are naturally uncertain and likely
overestimated due to Eddington bias \citep{morrissey07}. The detection
repeatability is generally low at the survey limit, and the detectability of
sources sometimes depends on subtle changes in the data analysis. For example,
two quasars that \citet{syphers09a} confirmed to show flux at \ion{He}{2}
Ly$\alpha$ were listed in the GR1 catalog, but not in further GALEX data
releases with improvements in survey depth, calibration and source detection
routines. While low-S/N detections might still indicate UV-transparent quasars,
we limit our statistical studies (\S\ref{sdssgalex}) to sources with S/N$>5$
in at least one of the GALEX bands. At the lowest S/N ratios encountered one
has to question the reality of the UV detection, in particular if a source
is seen just in one GALEX band. Sources formally detected in both bands should
be less affected, as source detection is performed independently on the FUV
and NUV images \citep{morrissey07}. Compared to the general incompleteness at
faint magnitudes, the subtle effect of PSF and sensitivity degradation at the
rim of the GALEX field of view can be neglected. We therefore performed our
correlation analysis on the full GALEX tiles, thereby maximizing the number
of promising UV-bright quasars for \ion{He}{2} studies.

\begin{figure}
\includegraphics[width=\columnwidth]{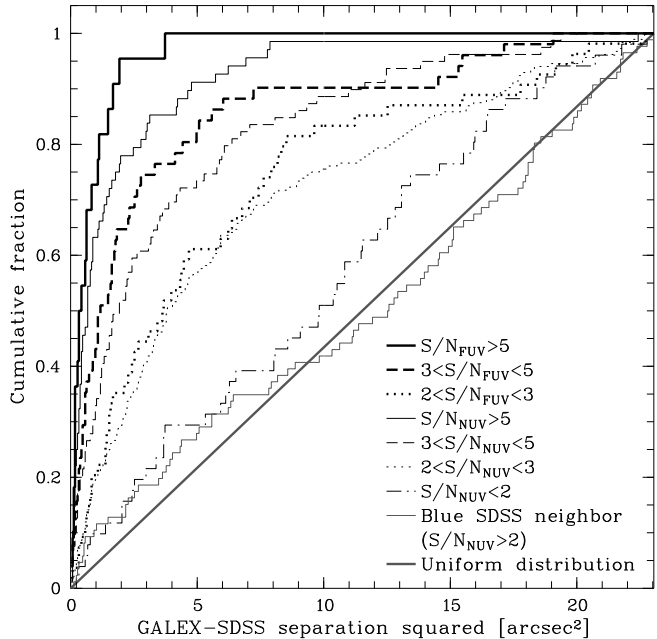}
\caption{\label{galexast}Cumulative fraction of the squared separation between the
GALEX positions and the optical position of GALEX-detected SDSS quasars. Thick (thin)
black lines show FUV-optical (NUV-optical) distributions for various ranges in S/N.
The thin gray line shows the cumulative distribution of the NUV-optical separations
of SDSS quasars having blue optical neighbors within $\simeq$5\arcsec.
The diagonal line denotes the uniform distribution with squared separation that is
expected for false positives.}
\end{figure}

We investigated the astrometric performance of GALEX in the low S/N regime by
calculating the offset between the optical quasar catalog position and the
GALEX NUV and/or FUV position. Given the nested GALEX surveys with a large
spread in depth, the astrometric accuracy primarily depends on S/N rather than
on magnitude. Figure~\ref{galexast} plots the cumulative fraction of the
squared separation between the GALEX positions and the optical position of
GALEX-detected SDSS $z_\mathrm{em}>2.7$ quasars for various ranges in S/N.
In this metric, false positives will be uniformly distributed in $r^2$,
whereas quasar (neighbor) matches should be concentrated at small (large)
offsets. Indeed, for SDSS quasars having blue optical neighbors within
5\arcsec, the distribution has two peaks, one at small separations for matches
to the quasar, and one at large separations corresponding to the detected blue
neighbor instead of the quasar. Therefore, it is essential to flag such cases
of potential source confusion caused by the broad GALEX PSF. With the assumption
that all GALEX sources in the SDSS footprint should have SDSS counterparts,
the GALEX sources without sufficiently blue optical neighbors are either UV
counterparts to the quasars in our catalog or false positives (noise).

Figure~\ref{galexast} shows that for SDSS quasars without blue optical neighbors
the distributions peak at small offsets with a clear dependence on S/N.
Almost all FUV (NUV) S/N$>5$ detections are within $r\la 2\arcsec$
($r\la 3\arcsec$) of the optical position with the difference being due to the
better resolution in the FUV \citep{morrissey07}. At lower S/N the astrometric
accuracy degrades and the rate of false positives should increase.
At S/N$_\mathrm{NUV}<2$ the cumulative fraction begins to resemble the one
expected for false positives, with the excess indicating some real detections
among them. Since the offset distributions at S/N$_\mathrm{NUV}>2$ are much more
concentrated, we infer that a limiting S/N$>2$ rather than a fixed limit in the
matching radius yields a source catalog of high purity and completeness. Our
chosen matching radius of $4.8\arcsec$ likely encompasses all true matches with
S/N$>3$, whereas a few real $2<$S/N$<3$ detections (without neighbors) might
exist at even larger separations. After excluding 117 ($\simeq 20$\%) of the
SDSS quasars with neighbors, restricting our catalog to S/N$>2$ (S/N$>3$)
in at least one GALEX band reduces the number of potential (probable) detections
to 601 (304).

We examined the GALEX source counts within 3\arcmin\ around our quasars to estimate
the probability of residual false matches between quasars and GALEX detections.
Despite their low resolution, GALEX images are confusion-limited only in the
longest DIS exposures \citep{hammer10} due to the low source density in the UV.
The measured density of S/N$_\mathrm{FUV}>2$ detections in a typical MIS exposure
is $\sim 1$/arcmin$^2$, which accounts for both real sources
\footnote{We compared our measured source density to the literature
\citep{bianchi07,hammer10}. At our low S/N threshold we only recover $\sim 60\%$
of the predicted sources on a given GALEX plate due to incompleteness at the
survey limit.} and false positives.
At this low of a source density, the chance for any S/N$_\mathrm{FUV}>2$ detection
to fall in our 4.8\arcsec\ aperture is small ($\la 2\%$). Given that the source
density on AIS plates is even lower, we conclude that essentially all FUV matches
on AIS and MIS plates will correspond to optical sources within the chosen aperture.
The rejection of SDSS quasars with blue neighbors probably excluded several real SDSS
quasar matches (Fig.~\ref{galexast}), so that we consider $\ga 98\%$ of the remaining
FUV-SDSS matches to be real. For non-SDSS quasars the remaining source confusion is
more important than the rate of spurious detections. Adopting our SDSS neighbor
fraction of $\sim 20\%$, we estimate a purity of $\sim 80\%$ for the quasars not
imaged by SDSS. Due to the challenging reduction and analysis of DIS plates,
we flagged the 23 quasars detected on DIS plates as still potentially affected by
source confusion (only 7 are in the constrained sample discussed in
\S\ref{sect_finalsample}).

\subsection{Comparison to Source Matching in \citet{syphers09b}}

\begin{figure}
\includegraphics[width=\columnwidth]{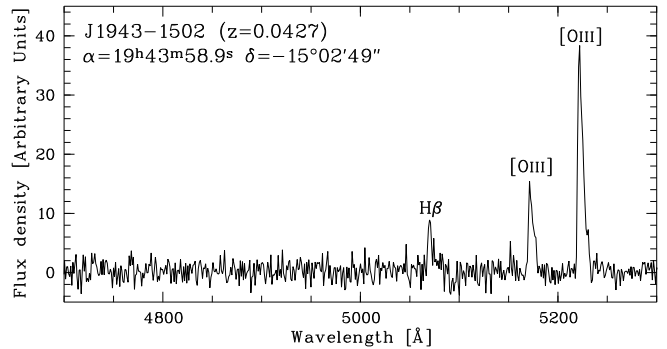}
\caption{\label{sypherscand}Lick/Kast spectrum of the emission line galaxy
J1943$-$1502 ($z=0.0427$). Identified emission lines are marked.}
\end{figure}

Recently, \citet{syphers09b} published a catalog of 593 sources detected in
GALEX GR4 and its small extension GR5. Apart from a slightly higher redshift
cutoff ($z>2.78$) and a smaller matching radius (3\arcsec\ around the quasar),
their approach to source matching (not target selection) was similar to ours.
However, they admitted that they did not verify the redshifts of the 165 sources
with GALEX GR4+5 counterparts stemming from the \citet{veron06} catalog.
\citet{syphers09b} presented follow-up HST/ACS UV prism spectroscopy of one of
these, J1943$-$1502, with an estimated slitless spectroscopic redshift of 3.3
\citep{crampton97}. In order to establish whether this object can be used for
\ion{He}{2} IGM studies, we obtained an optical spectrum with the Kast
spectrograph at the 3-m Shane Telescope at Lick Observatory. We confirm
J1943$-$1502 as a naturally UV-bright low-redshift emission line galaxy rather
than a quasar (Fig.~\ref{sypherscand}). We caution that the \citet{syphers09b} list of
\citet{veron06} sources contains 41 more such candidates the redshifts of which should be
confirmed before embarking on follow-up UV spectroscopy with HST. In addition, 5 other
sources from the \citet{veron06} catalog that are listed by \citet{syphers09b} as
GALEX-detected $z_\mathrm{em}>2.7$ quasars are actually at lower redshifts according
to our visual inspection of their spectra.

\begin{figure*}
\includegraphics[width=\textwidth]{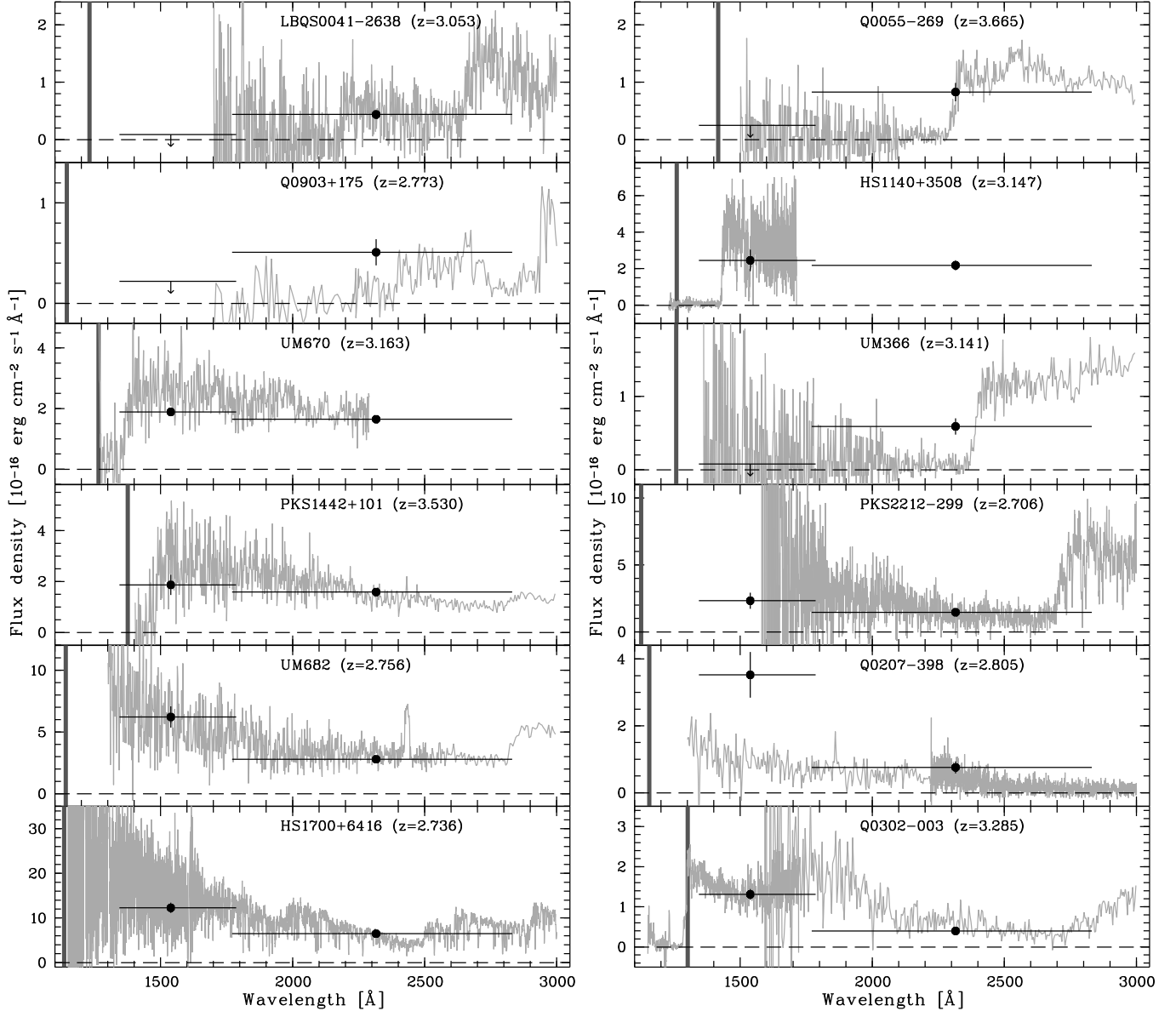}
\caption{\label{hstspc}HST spectra of 12 quasars (gray) and NUV \& FUV fluxes 
measured by GALEX (filled circles with error bars and indicated bandpass).
Arrows indicate upper limits from GALEX non-detections. Dashed lines mark zero
flux. The thick vertical lines indicate the expected onset of \ion{He}{2} absorption.}
\end{figure*}

\subsection{The GALEX UV colors of high-redshift quasars}
\label{obscolors}

The large sky coverage of GALEX enables the recovery of many UV-bright
$z_\mathrm{em}>2.7$ quasars that have previously been followed up with HST to
search for \ion{He}{2} absorption by the IGM. GALEX recovers all 8
quasars known to show flux at \ion{He}{2} Ly$\alpha$ that had been selected for
observations before the launch of GALEX. \citet{syphers09b,syphers09a} recently
confirmed 22 GALEX-selected sightlines to show \ion{He}{2}, and all but the two
listed only in GR1 are contained in the GALEX GR4 source catalog. 13 of the total
30 confirmed \ion{He}{2} quasars are detected by GALEX at a low S/N$<3$, and we
suspect that there is a larger population of UV-transparent quasars missed at the
GALEX survey limit. We also recovered UV-bright quasars considered in previous
photometric and spectroscopic surveys for \ion{He}{2} with HST, the sightlines of
which are intercepted by optically thick Lyman limit systems redward of the onset
of \ion{He}{2} absorption.

\tabletypesize{\footnotesize}
\begin{deluxetable*}{lcrrlcl}
\tablecaption{\label{hstlist}Data on the UV-bright quasars shown in Fig.~\ref{hstspc}}
\tablewidth{0pt}
\tablehead{
\colhead{Object}&\colhead{$z_\mathrm{em}$}&\colhead{$m_\mathrm{FUV}$\,[AB]}&\colhead{$m_\mathrm{NUV}$\,[AB]}&\colhead{HST spectrum}
&\colhead{$z_\mathrm{LLS}$}&\colhead{References}
}
\startdata
PKS~2212$-$299	&$2.706$	&$20.74$	&$20.35$	&STIS G230L		&$0.6329$	&\citet{rao06}\\
HS~1700$+$6416	&$2.736$	&$18.94$	&$18.74$	&FOS G130H/G190H/G270H	&\nodata	&\citet{reimers92,evans04}\\
UM~682		&$2.756$	&$19.68$	&$19.65$	&FOS G160L/PRISM	&\nodata	&HST Archive\\
Q~0903$+$175	&$2.773$	&$>23.32$	&$21.50$	&FOS G160L/G270H	&BAL		&\citet{turnshek96}\\
Q~0207$-$398	&$2.805$	&$20.29$	&$21.07$	&FOS G160L/G270H	&\nodata	&\citet{bechtold02}\\
LBQS~0041$-$2638&$3.053$	&$>24.32$	&$21.67$	&STIS G230L		&$1.38:$	&HST Archive\\
UM~366		&$3.141$	&$>24.40$	&$21.35$	&FOS G160L/G270H	&$1.6128$	&\citet{rao00,evans04}\\
HS~1140+3508	&$3.147$	&$20.68$	&$19.92$	&STIS G140L		&$0.557$	&HST Archive\\
UM~670		&$3.163$	&$20.97$	&$20.22$	&FOS G160L		&$0.47:$	&\citet{lyons94,evans04}\\
Q~0302$-$003	&$3.285$	&$21.37$	&$21.76$	&STIS G140L/G230L	&\nodata	&\citet{jakobsen94,heap00}\\
PKS~1442$+$101	&$3.530$	&$20.98$	&$20.26$	&FOS G160L/PRISM	&$0.621:$	&\citet{lyons95,evans04}\\
Q~0055$-$269	&$3.665$	&$>23.17$	&$20.97$	&FOS G160L/PRISM	&$1.5335$	&\citet{cristiani95,evans04}
\enddata
\end{deluxetable*}

In Fig.~\ref{hstspc} we compare the GALEX fluxes of 12 quasars to their UV
spectra taken with HST. Their GALEX UV magnitudes are provided in
Table~\ref{hstlist} together with references to the UV spectra and the Lyman
limit systems zeroing the spectral flux (if any). As these quasars are bright in
the UV they are imaged with GALEX at high S/N, so that the GALEX fluxes are in
very good agreement with the HST spectrophotometry. More interestingly, we find
that several opaque sightlines are just detected in the NUV, but not in the FUV
as expected (LBQS~0041$-$2603, Q~0055$-$269 and UM~366 in Fig.~\ref{hstspc}).
In contrast, quasars that show flux down to the onset of \ion{He}{2} absorption
are detected in both bands with the flux rising towards shorter wavelengths as
it recovers from partial Lyman limit systems (HS~1700$+$6416 and Q~0302$-$003).
Thus, the GALEX UV color $m_\mathrm{FUV}-m_\mathrm{NUV}$ can be used to
efficiently distinguish between opaque sightlines (red UV color) and transparent
ones (blue UV color). The only quasars that remain insensitive to this obvious
color selection criterion are those caught by an optically thick Lyman limit
break just in the narrow range between the GALEX FUV band and the onset of
\ion{He}{2} absorption (HS~1140$+$3508, UM~670, PKS~1442$+$101, PKS~2212$-$299
in Fig.~\ref{hstspc}). We also identify two FUV-detected quasars, the HST
spectra of which do not extend to \ion{He}{2} Ly$\alpha$ in the rest frame of
the quasar, located near the UV sensitivity cutoff of HST (UM~682 and Q~0207$-$398).
These two sightlines are likely transparent, as there are no obvious strong
Ly$\alpha$ absorbers that could cause a Lyman limit break in the $\sim 200$\AA\
gap to the onset of \ion{He}{2} absorption. 

With the additional quasars targeted in recent surveys for \ion{He}{2} sightlines
\citep{syphers09b,syphers09a} we can confirm the trend that most quasars with
flux down to \ion{He}{2} Ly$\alpha$ show blue GALEX colors, whereas most
fruitlessly targeted quasars are characterized by red colors
(see Fig.~\ref{gr4zcolor} below). Although more uncertain at low S/N, the colors
still distinguish both quasar populations at S/N$\ga 3$. Excluding sources with
neighbors, $\sim 50$\% of the SDSS quasars in our sample are detected at S/N$>3$
in the NUV band, but are lacking a significant FUV detection (S/N$_\mathrm{FUV}<2$),
indicating the ubiquitous strong Lyman continuum absorption. In particular, FUV
dropouts detected in the NUV at high significance likely correspond to optically thick
Lyman limit breaks.
 
In the following sections we will further explore how to further constrain our sample
by the GALEX UV color to select the most promising quasar sightlines to detect
\ion{He}{2} absorption. This requires one to create mock quasar spectra with
appropriate \ion{H}{1} absorption, and to perform GALEX photometry on them to relate
the GALEX UV color to the Lyman continuum absorption along the line of sight.

\section{Monte Carlo simulations of high-redshift quasar spectra}
\label{mcspectra}

\subsection{Monte Carlo model for the \ion{H}{1} Lyman series and Lyman continuum absorption}
\subsubsection{General procedure}
\label{model_gen}

For the problem at hand we followed standard practice to generate Monte Carlo
(MC) \ion{H}{1} Lyman forest and Lyman continuum absorption spectra from the
observed statistical properties of the Ly$\alpha$ forest
\citep[e.g.][]{moller90,madau95,bershady99,inoue08}. The spectra were generated
under the null hypothesis that the Ly$\alpha$ forest can be approximated as a
random collection of absorption lines (Voigt profiles) with uncorrelated
parameters (redshift $z$, column density $N_\mathrm{H\,I}$ and Doppler parameter $b$).
From the line list representing the \ion{H}{1} absorber population on a given
line of sight from $z=0$ to an emission redshift $z_\mathrm{em}$ we created
absorption spectra of the Lyman series (up to Ly30). Individual resolved Voigt
profiles were computed on $\Delta\lambda=0.05$\AA\ pixels using the
approximation by \citet{teppergarcia06}. Lyman continuum absorption was
included using the \ion{H}{1} ionization cross section by \citet{verner96}.

In order to accurately predict the far-UV attenuation of high-redshift quasars
by the IGM we desired a model that successfully reproduces the observed
statistical properties of the Ly$\alpha$ forest at all redshifts, in particular
concerning high-column density absorbers. Considering the recent observational
advances in Ly$\alpha$ forest statistics, we deviated from previous simple MC
descriptions of the Ly$\alpha$ forest and adjusted our input parameters as
detailed in the following.

\subsubsection{The absorber redshift distribution function}
\label{model_z}

In our MC model the number of \ion{H}{1} absorbers per line of sight in a given
redshift range is a Poisson process \citep{zuophinney93}. The observed mean
differential line density per unit redshift is commonly parameterized as a
power law $\mathrm{d}n/\mathrm{d}z|_\mathrm{forest}\propto (1+z)^\gamma$ that
results in an effective optical depth
$\tau_\mathrm{eff,\alpha}\propto (1+z)^{\gamma+1}$ for Ly$\alpha$
(and higher order series) absorption \citep{zuo93}. While there is some evidence
that the redshift evolution depends on the column density even in the low-column
density Ly$\alpha$ forest, the uncertainties are still large due to the
non-unique process to deblend the forest into a series of Voigt profiles
especially at $z\ga 3$, incompleteness at the lowest column densities
($\log N_\mathrm{H\,I}\la 12.5$), and the paucity of moderate-column density
($\log N_\mathrm{H\,I}\ga 14.5$) systems \citep{kim97,kim02}. We therefore chose
to parameterize $\mathrm{d}n/\mathrm{d}z$ for absorbers with
$12<\log N_\mathrm{H\,I}<19$ as a single power law, the parameters of which were
fixed by requiring each simulated spectrum to be consistent with a specified
power law in $\tau_\mathrm{eff,\alpha}(z)$. Observations point to a break at
$z\sim 1.5$, below which there is little evolution both in the line density
\citep[e.g.][]{weymann98,kim02,janknecht06} and the mean absorption in the
Ly$\alpha$ forest $D_\mathrm{A}=1-\mathrm{e}^{-\tau_\mathrm{eff,\alpha}}$
\citep{kirkman07}. Thus, we assumed a broken power law for
$\tau_\mathrm{eff,\alpha}(z)$. Knowing that a power-law line distribution
generally will not yield a power law for $D_\mathrm{A}(z)$ assumed by
\citet{kirkman07}, we converted their $D_\mathrm{A}$ to
$\tau_\mathrm{eff,\alpha}$ and obtained a fit
$\tau_\mathrm{eff,\alpha}(z)=0.017\left(1+z\right)^{1.20}$ for $z\la 1.6$.
At $2\la z\la 4$ $\tau_\mathrm{eff,\alpha}$ has been precisely measured in
high-resolution spectra \citep{kim07b,faucher08,dallaglio08}, and the remaining
disagreement at $z\ga 4$ is likely due to continuum uncertainties, where very
few pixels remain unabsorbed even in high resolution spectra. We adopted the
fit $\tau_\mathrm{eff,\alpha}=0.0062\left(1+z\right)^{3.04}$ from
\citet{dallaglio08}, valid at $1.8<z<4.6$. Note that the break redshift cannot
be determined as the intersection of the two power laws, since this would
require one to extrapolate $\tau_\mathrm{eff,\alpha}(z)$ beyond the quoted
validity ranges. Since the break is observationally not well constrained,
given the large scatter of $\tau_\mathrm{eff,\alpha}$ measurements at $1.7<z<2$
and the paucity of data at $z\sim 1.5$ \citep{kirkman07}, we adopted a break
redshift of $z=1.5$ for the broken power law in $\tau_\mathrm{eff,\alpha}(z)$.

\begin{figure}
\includegraphics[width=\columnwidth]{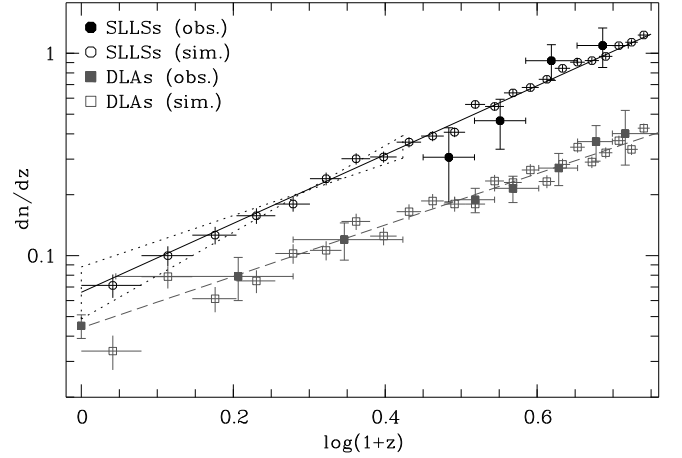}
\caption{\label{simsllsdla}Adopted differential number densities
$\mathrm{d}n/\mathrm{d}z$ of SLLSs and DLAs as a function of redshift $z$. Filled
symbols represent observed data (see text), whereas open symbols show the
distributions recovered from 4000 MC simulations. The straight solid and dashed
lines denote the power law fits to the observed data adopted for the simulations.
The dotted lines show different normalizations to yield the number density of
SLLSs at $z<1.65$ observed by \citet{rao06} adopting the slope of the DLA
evolution law and the SLLS slope found at high $z$.}
\end{figure}

For $\log N_\mathrm{H\,I}\ge 19$ absorbers we had to assume different redshift
evolution laws, both because these systems are generally excluded in fits of
$\tau_\mathrm{eff,\alpha}(z)$, and due to the fact that their number densities
seem to evolve much slower with redshift. For damped Ly$\alpha$ systems
(DLAs, $\log N_\mathrm{H\,I}\ge 20.3$) we adopted
$\mathrm{d}n/\mathrm{d}z|_\mathrm{DLA}=0.044\left(1+z\right)^{1.27}$, determined
by \citet{rao06} over the redshift range $0<z<5$. Figure~\ref{simsllsdla}
compares the observed number densities of DLAs compiled by \citet{rao06} to mock
number densities obtained on 4000 MC sightlines assuming their fit for
$\mathrm{d}n/\mathrm{d}z|_\mathrm{DLA}$. For Super Lyman Limit systems
(SLLSs, $19\le\log N_\mathrm{H\,I}<20.3$) there are significantly less
constraints in the literature. A maximum-likelihood power-law fit to the SLLS
survey by \citet{omeara07} yields
$\mathrm{d}n/\mathrm{d}z|_\mathrm{SLLS}=0.034\left(1+z\right)^{2.14}$ at
$1.8<z<4.2$, but extrapolation to lower redshifts underestimates the lower limit
$n_\mathrm{SLLS}\ga 2n_\mathrm{DLA}$ at $z<1.65$ given by \citet{rao06}. Rather
than a break in the number density of SLLSs, this probably indicates that a much
larger redshift range is needed to accurately describe the number density
evolution of the rare SLLSs. By constraining the slope to
$1.27<\gamma_\mathrm{SLLS}<2.14$ (i.e. between the evolution rate of DLAs and
the SLLS fit at high $z$), and considering the estimated total number of
$z<1.65$ SLLSs by \citet{rao06} we obtained a rough constraint on the
low-redshift evolution of SLLSs (dotted lines in Fig.~\ref{simsllsdla}). After
binning the high-$z$ measurements by \citet{omeara07} we determined
$\mathrm{d}n/\mathrm{d}z|_\mathrm{SLLS}\simeq 0.066\left(1+z\right)^{1.70}$ by
eye (Fig.~\ref{simsllsdla}), noting that these numbers are quite uncertain as
the $z<2$ SLLS population is not well constrained.

\subsubsection{The Doppler parameter distribution function}

Although the Doppler parameter distribution function $\mathrm{d}n/\mathrm{d}b$
is not required to calculate the attenuation of quasars by the IGM below the
Lyman limit, our MC simulations reproduce the observed effective optical depth
in the Ly$\alpha$ forest instead of a line density distribution. As the
equivalent width of the lines on the flat part of the curve of growth
($13.5\la\log N_\mathrm{H\,I}\la 18.5$) depends both on the column density
$N_\mathrm{H\,I}$ and the Doppler parameter $b$, the line density that is
consistent with our adopted $\tau_\mathrm{eff,\alpha}(z)$ implicitly depends
on the Doppler parameter distribution. For simplicity, we adopted the single
parameter distribution function suggested by \citet{hui99},
$\mathrm{d}n/\mathrm{d}b\propto b^{-5}\exp\left(-b^4/b_\sigma^4\right)$, with
$b_\sigma=24\,\mathrm{km}\,\mathrm{s}^{-1}$ \citep{kim01} independent of redshift
and column density, and restricted to the plausible range
$10\,\mathrm{km}\,\mathrm{s}^{-1}\le b<100\,\mathrm{km}\,\mathrm{s}^{-1}$.

\subsubsection{The column density distribution function}
\label{model_cddf}

Previous studies on the IGM attenuation of high-redshift sources approximated
the column density distribution function (CDDF) by a single or a broken power
law, mainly driven by the reasonable approximation of the CDDF as a single
power law $\mathrm{d}n/\mathrm{d}N_\mathrm{H\,I}\propto N_\mathrm{H\,I}^{-\beta}$
with $\beta\simeq 1.5$ over practically the full observable column density range
\citep{tytler87}. However, more recent studies revealed significant deviations
in the high-redshift CDDF from a single power law at intermediate
\citep[$14.5\la\log N_\mathrm{H\,I}\la 16$,][]{petitjean93,hu95,kim02} and at
the highest column densities
\citep[$\log N_\mathrm{H\,I}\ga 19$,][]{storrie00,prochaska05,omeara07}.
A careful treatment of these systems is necessary, since even the intermediate
column densities have a strong impact on the total LyC absorption
\citep{madau95,haardt96}. However, due to the scarcity of
$14.5\la\log N_\mathrm{H\,I}\la 17$ systems, the shape of the CDDF in this
important range is presently not well constrained \citep{kim02}.

We took a novel approach to constrain the high-$z$ CDDF at
$14.5<\log N_\mathrm{H\,I}<19$ by matching the mean free path (MFP) to Lyman
limit photons calculated from the CDDF to our recent measurements from SDSS at
$3.6<z<4.2$ \citep[][see also \citealt{prochaska10}]{prochaska09}. The
effective optical depth to Lyman limit photons emitted at $z_\mathrm{em}$ and
observed at $z_\mathrm{obs}$ is \citep[e.g.][]{paresce80}
\begin{eqnarray}\label{eq_mfp}\nonumber
\tau_\mathrm{eff,LL}\left(z_\mathrm{obs},z_\mathrm{em}\right)&=&\int_{z_\mathrm{obs}}^{z_\mathrm{em}}\int_0^\infty f\left(N_\mathrm{H\,I},z\right)\\
&&\times\left[1-\mathrm{e}^{-N_\mathrm{H\,I}\sigma_\mathrm{LL}\left(\frac{1+z}{1+z_\mathrm{obs}}\right)^{-3}}\right]\,\mathrm{d}N_\mathrm{H\,I}\mathrm{d}z,
\end{eqnarray}
with the Lyman limit photoionization cross section
$\sigma_\mathrm{LL}=6.33\times 10^{-18}\mathrm{cm}^2$ and the frequency
distribution of absorbers in redshift and column density
$f\left(N_\mathrm{H\,I},z\right)=\frac{\partial^2 n}{\partial N_\mathrm{H\,I}\partial z}$.
Considering the different power-law redshift distributions of different absorber
populations as outlined above, we approximated the CDDF as piecewise power laws
that do not change over the considered redshift range, yielding
\begin{eqnarray}\label{eq_taull}\nonumber
\tau_\mathrm{eff,LL}\left(z_\mathrm{obs},z_\mathrm{em}\right)&=&\sum_j C_j \int_{z_\mathrm{obs}}^{z_\mathrm{em}}\int_{N_{\mathrm{H\,I,min,j}}}^{N_{\mathrm{H\,I,max},j}}\left(1+z\right)^{\gamma_j}N_\mathrm{H\,I}^{-\beta_j}\\
&&\times\left[1-\mathrm{e}^{-N_\mathrm{H\,I}\sigma_\mathrm{LL}\left(\frac{1+z}{1+z_\mathrm{obs}}\right)^{-3}}\right]\,\mathrm{d}N_\mathrm{H\,I}\mathrm{d}z,
\end{eqnarray}
with different normalizations $C_j$ and power law exponents
$\left(\gamma_j,\beta_j\right)$ in different column density ranges
$\left[N_{\mathrm{H\,I,min},j},N_{\mathrm{H\,I,max},j}\right]$.
The normalization constants $C_j$ are the products of the line density
normalizations $A_j$ ($\mathrm{d}n/\mathrm{d}z=A_j\left(1+z\right)^{\gamma_j}$)
and the CDDF normalizations to yield an integral of unity in the respective
column density range
\begin{equation}
C_j=\frac{A_j\left(1-\beta_j\right)}{N_{\mathrm{H\,I,max},j}^{1-\beta_j}-N_{\mathrm{H\,I,min},j}^{1-\beta_j}}.
\end{equation}
For the SLLSs we assumed $\beta_\mathrm{SLLS}=1.4$ \citep{omeara07}, whereas
for DLAs we adopted $\beta_\mathrm{DLA}=2$ \citep{prochaska05}. We fixed the
contributions of SLLSs and DLAs to $\tau_\mathrm{eff,LL}$ with our explicit
line density evolutions. These absorbers are highly optically thick to LyC
photons, so their incidence rather than their column density distribution
determines their share to $\tau_\mathrm{eff,LL}$.

By definition the MFP corresponds to the proper distance where
$\tau_\mathrm{eff,LL}=1$ for Lyman limit photons emitted at $z_\mathrm{em}$.
In order to constrain the shape of the CDDF of Lyman limit systems and the
Ly$\alpha$ forest, we considered a contiguous triple power law at
$12<\log N_\mathrm{H\,I}<19$ that results in a quasi-continuous CDDF over the
full column density range. Requiring the $12<\log N_\mathrm{H\,I}<19$ forest to
result in our assumed power-law redshift evolution of the effective Ly$\alpha$
optical depth, we fixed $\gamma=2.04$ ($\gamma=0.20$) for
$\log N_\mathrm{H\,I}<19$ at $z>1.5$ ($z\le 1.5$). We then varied the triple
power law CDDF, each time simulating 1000 MC sightlines at $0<z<4.6$ in order
to determine the normalization constants for the line densities $A_j$ followed
by computing the resulting total Lyman limit effective optical depth
(eq.~\ref{eq_taull}) and comparing the corresponding MFP at $3.6<z<4.2$ to our
measurements \citep{prochaska09}.

In order to find the most plausible values for the slopes and breaks in the
CDDF we considered additional observational constraints. The CDDF is best
determined in the $z\sim 3$ Ly$\alpha$ forest and we adopted $\beta_1=1.5$ for
$12<\log N_\mathrm{H\,I}<14.5$ at $z>1.5$ \citep[e.g.][]{hu95}. At
$\log N_\mathrm{H\,I,max,1}=14.5$ we imposed the first break in the CDDF to
account for the deficit of absorbers at $\log N_\mathrm{H\,I}\ga 14.5$
\citep[e.g.][]{petitjean93}. Initially, we tried a single $\beta=1.5$ power
law that strongly underpredicted the MFP, but remarkably extrapolates into the
SLLS and DLA range where the CDDF was set independently. This probably reflects
the fact that the $\beta=1.5$ power law approximation relies on both ends of
the CDDF, which are by far the best constrained. We then varied the second break
column density $N_\mathrm{H\,I,max,2}$ and the slope $\beta_2$ between the two
breaks, requiring the slope $\beta_3$ at
$\log N_\mathrm{H\,I,max,2}<\log N_\mathrm{H\,I}<19$ to meet the extrapolated
$\beta=1.5$ power law at $\log N_\mathrm{H\,I}=19$, thus yielding a
quasi-continuous CDDF, which we used at $z>1.5$.

\begin{figure}
\includegraphics[width=\columnwidth]{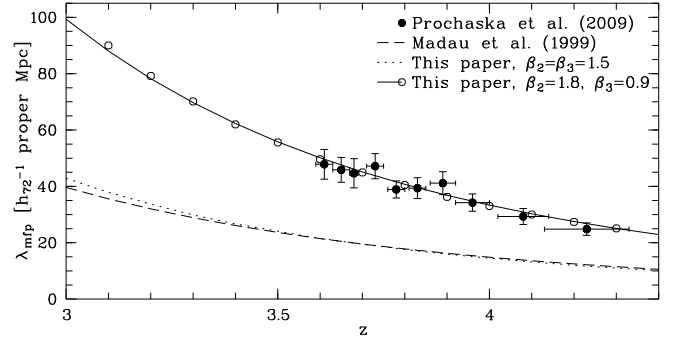}
\caption{\label{mfpmatch}Comparison of the mean free path $\lambda_\mathrm{mfp}$
as a function of redshift $z$ resulting from our adopted redshift evolution and
column density distribution of Ly$\alpha$ absorbers (open circles), direct
measurements from \citet[][filled circles]{prochaska09} and the previous
theoretical estimate by \citet{madau99}. All values are reported for a flat
cosmological model with $\left(\Omega_\mathrm{m},\Omega_\Lambda\right)=(0.3,0.7)$
and $H_0=72\,\mathrm{km}\,\mathrm{s}^{-1}\,\mathrm{Mpc}^{-1}$. The solid line
shows a power law fit to our empirical estimates
$\lambda_\mathrm{mfp}=50.14\left[\left(1+z\right)/4.6\right]^{-4.89}$ Mpc, valid
for $z>3$. The dotted line shows the mean free path implied by a single
$\beta=1.5$ power law at $\log N_\mathrm{H\,I}<19$ instead of the adopted broken
power law.}
\end{figure}

Our calculations confirmed previous results that the MFP, and thus the mean LyC
absorption at high redshift, is very sensitive to the shape of the CDDF at
intermediate column densities \citep[e.g.][]{madau95}. In particular, we could
rule out many parameter combinations
$\left(\log N_\mathrm{H\,I,max,2},\beta_2\right)$ by requiring the calculated
MFP to be consistent with both the normalization and the redshift evolution of
the measured MFP at $z>3.6$. In Fig.~\ref{mfpmatch} we show our best match to
the actual observations, obtained for
$\left(\log N_\mathrm{H\,I,max,2},\beta_2\right)=(17.5,1.8)$, which imply a
remarkably flat $\beta_3\simeq 0.9$. The modeled MFP agrees extremely well with
the observed values and can be accurately described by a power law at $z>3$,
yielding $\lambda_\mathrm{mfp}=50.14\left[\left(1+z\right)/4.6\right]^{-4.89}$
proper Mpc for a flat cosmology with
$\left(\Omega_\mathrm{m},\Omega_\Lambda\right)=(0.3,0.7)$ and
$H_0=72\,\mathrm{km}\,\mathrm{s}^{-1}\,\mathrm{Mpc}^{-1}$. In contrast, by
adopting a featureless $\beta=1.5$ power law at $\log N_\mathrm{H\,I}<19$
together with the the slightly different distributions for the higher column
density systems, the MFP is smaller by a factor $\simeq 2.3$ and is strongly
inconsistent with the MFP measurements. The very good agreement between this
underestimate and the MFP adopted by \citet{madau99} is not too surprising,
as they assumed a single $\beta=1.5$ and a single absorber population evolving
with redshift at $\gamma=2$, very similar to the $\gamma=2.04$ we adopted for
$\log N_\mathrm{H\,I}<19$. We emphasize that at least two inflections in the
CDDF are required at $\log N_\mathrm{H\,I}<19$ in order to yield a
quasi-continuous CDDF that is consistent with our direct MFP measurements
\citep[see also][]{prochaska10}.

\begin{figure}
\includegraphics[width=\columnwidth]{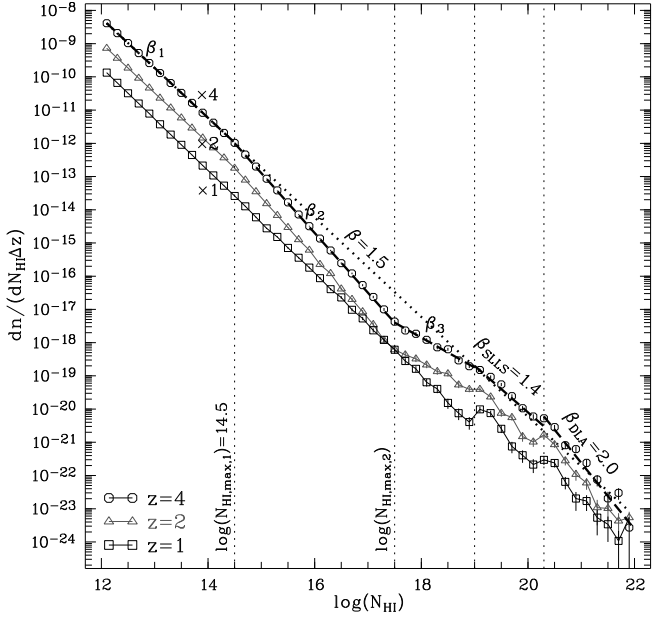}
\caption{\label{modelcddf}Modeled column density distribution functions (CDDFs)
$\mathrm{d}n/\left(\mathrm{d}N_\mathrm{H\,I}\Delta z\right)$ as a function of
\ion{H}{1} column density $N_\mathrm{H\,I}$ in a range $\Delta z=0.1$ around
$z=1$, $z=2$ and $z=4$. For clarity the CDDFs have been scaled by the indicated
factors. The symbols represent the binned CDDFs recovered from 4000 MC sightlines.
The thick dashed lines show fits to the column density distribution at $z=4$ with
the different slopes adopted in different column density regions (vertical dotted
lines) designed to yield the measured MFP and its redshift evolution. The thick
dotted line shows the CDDF at $z=4$ adopted for the forest ($\beta=1.5$)
extrapolated to high column densities. Note that the redshift distributions and
CDDFs in the SLLS and DLA range are set independently, whereas continuity is
required for $\log N_\mathrm{H\,I}<19$.}
\end{figure}

Figure~\ref{modelcddf} shows the corresponding model CDDFs at $z=4$
(covered by our MFP measurements) and $z=2$ (extrapolated from higher redshifts
using the redshift evolution laws from Section~\ref{model_z}). The CDDF at
$\log N_\mathrm{H\,I}>19$ is remarkably smooth, given that independent and
uncertain redshift evolution laws set the CDDF normalization there. The
requirement for the CDDF to match the $\beta=1.5$ power law extrapolation from
the low column density forest at $\log N_\mathrm{H\,I}\simeq19$ yields a
continuous CDDF, both at $z=4$ and at $z=2$, as intended.

\begin{figure}
\includegraphics[width=\columnwidth]{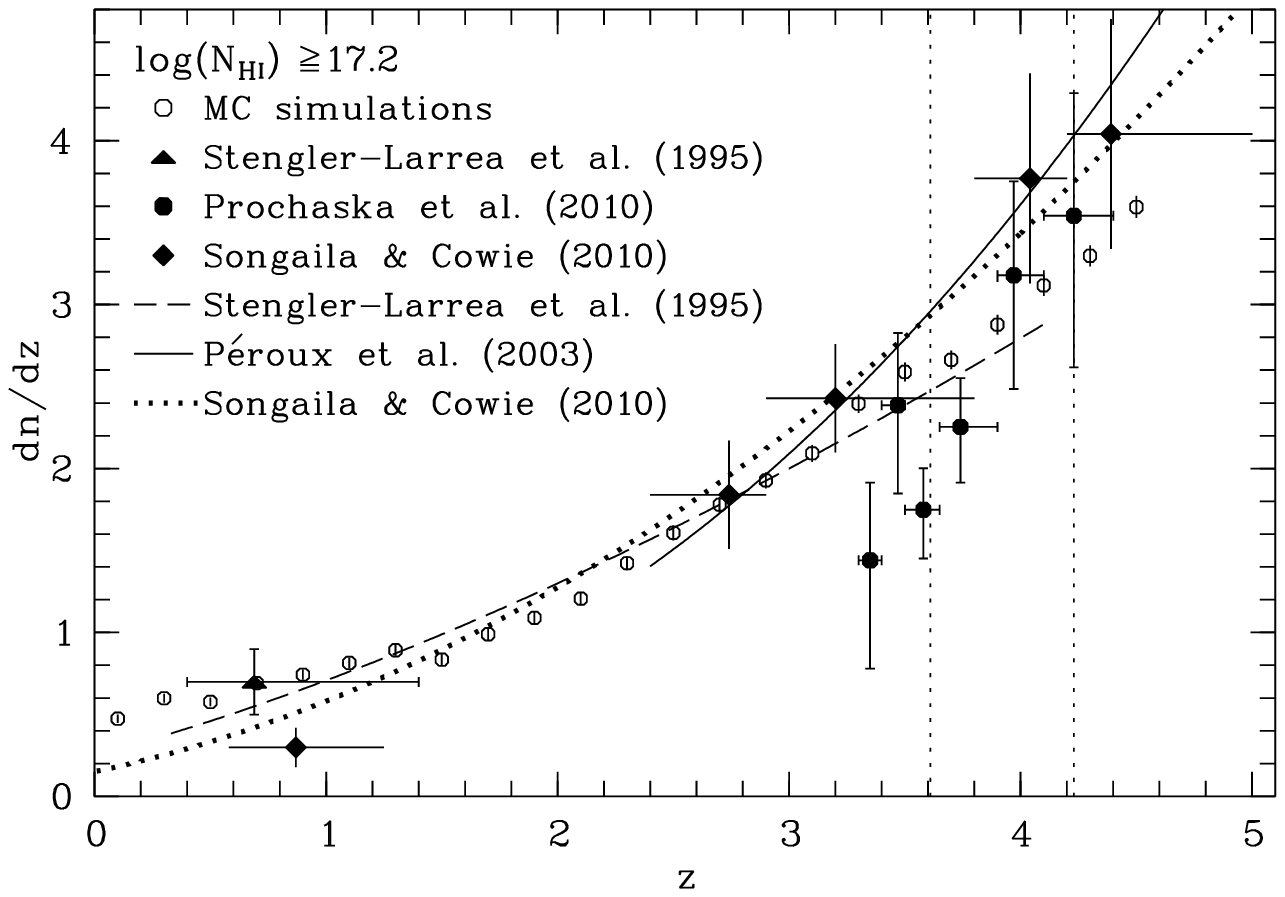}
\caption{\label{dndz_logNge17p2}
Differential number density $\mathrm{d}n/\mathrm{d}z$ for Lyman limit systems
($\log{N_\mathrm{H\,I}}\ge 17.2$) as a function of redshift $z$ predicted from our
MC simulations (open circles) compared to power law fits to actual data from
\citet{stengler-larrea95} (dashed), \citet{peroux03} (solid) and \citet{songaila10}
(thick dotted) in their quoted validity range. The filled symbols show actual
measurements \citep{stengler-larrea95,prochaska10,songaila10}. Note that the fit
from \citet{songaila10} includes previous data from \citet{stengler-larrea95} at
low $z$ and \citet{peroux03} at high $z$. The vertical dotted lines mark the redshift
range of our MFP measurements which constrain the number of Lyman limit systems
in the MC simulations.}
\end{figure}

As a consistency check we used our chosen distribution parameters to predict the
incidence of Lyman limit systems (LLSs; $\log N_\mathrm{H\,I}\ge 17.2$).
Figure~\ref{dndz_logNge17p2} compares the mock differential number densities of
LLSs to observations based on line counting
\citep{stengler-larrea95,peroux03,prochaska10,songaila10}. Given the large statistical and
systematic uncertainties in the observations, the agreement is remarkable, even
at $1.5<z<3.6$, where we rely on the CDDF extrapolation from higher redshifts.
While our MFP measurements tightly constrain the incidence of LLSs at $z>3.6$,
the extrapolated CDDF might underestimate the incidence of LLSs if the CDDF
straightens at lower $z$. By the same token, if LLSs evolve as strongly as
indicated by \citet{prochaska10}, we might have underestimated the MFP at $z<3.6$.
Our prediction for the evolution of LLSs is most consistent with the fit by
\citet{stengler-larrea95}, who sampled $z\sim 3$ based on earlier studies. Better
models and predictions hinge on measurements of the MFP and the incidence of
LLSs at $z\simeq 2$--3.

At low redshifts the CDDF is considerably less constrained, as the declining
line density requires large samples of sightlines to be observed from space.
\citet{janknecht06} determined a single power law for the CDDF at $0.5<z<1.9$
with $\beta=1.6$, but their fit is dominated by the low column density forest
and slightly overpredicts the fraction of $\log N_\mathrm{H\,I}\ga 14.5$
lines (their Fig.~5). \citet{lehner07} found that the $z<0.4$ CDDF steepens
further at low column densities, whereas $\log N_\mathrm{H\,I}\ga 14.5$ lines
show a flatter slope $\beta\sim 1.5$. The low-redshift observations are
inconsistent with our high-$z$ model CDDF with its inferred low abundance of
$14.5\la\log N_\mathrm{H\,I}\la 17.5$ absorbers. For simplicity, we therefore
assumed a featureless power law at $z<1.5$ for the column density range
$12<\log N_\mathrm{H\,I}<19$, the slope of which was constrained by requiring
a rough match to the CDDF at $\log N_\mathrm{H\,I}\approx 19$
(set independently by the SLLS distribution from above), while yielding the
observed number of LLSs at low redshifts \citep{stengler-larrea95} and
preserving the continuity in $\mathrm{d}n/\mathrm{d}z$ for LLSs predicted
from our extrapolation from higher redshifts (Fig.~\ref{dndz_logNge17p2}).
A slope $\beta=1.55$ matched these requirements. As an example, we show the
modeled $z=1$ CDDF in Fig.~\ref{modelcddf}. A lower incidence of LLSs at $z\sim 1$
as recently indicated by \citet{songaila10} would not drastically change our
predictions, because the total LyC absorption at the \ion{He}{2} edge primarily
depends on the sparsely sampled redshift range $z\simeq 2$--3.

\tabletypesize{\footnotesize}
\begin{deluxetable}{ccccccc}
\tablewidth{\columnwidth}
\setlength{\tabcolsep}{0.015in}
\tablecaption{\label{simpar}Monte Carlo simulation parameters}
\tablehead{
\colhead{$z$ range}&\colhead{$z$ norm.\tablenotemark{a}}&\colhead{$\gamma$}&\colhead{$N_\mathrm{H\,I}$ range}&\colhead{$\beta$\tablenotemark{b}}&\colhead{$b_\sigma$\tablenotemark{c}}&\colhead{$b$ range}\\
\colhead{}&\colhead{}&\colhead{}&\colhead{[cm$^{-2}$]}&\colhead{}&\multicolumn{2}{c}{[km$\,$s$^{-1}$]}
}
\startdata
$\left[0.0,1.5\right]$	&$B=0.0170$	&$0.20$	&$\left[10^{12.0},10^{19.0}\right)$	&$1.55$	&$24$	&$\left[10,100\right)$\\
$\left(1.5,4.6\right]$	&$B=0.0062$	&$2.04$	&$\left[10^{12.0},10^{14.5}\right)$	&$1.50$	&$24$	&$\left[10,100\right)$\\
$\left(1.5,4.6\right]$	&$B=0.0062$	&$2.04$	&$\left[10^{14.5},10^{17.5}\right)$	&$1.80$	&$24$	&$\left[10,100\right)$\\
$\left(1.5,4.6\right]$	&$B=0.0062$	&$2.04$	&$\left[10^{17.5},10^{19.0}\right)$	&$0.90$	&$24$	&$\left[10,100\right)$\\
$\left[0.0,4.6\right]$	&$A=0.0660$	&$1.70$	&$\left[10^{19.0},10^{20.3}\right)$	&$1.40$	&$24$	&$\left[10,100\right)$\\
$\left[0.0,4.6\right]$	&$A=0.0440$	&$1.27$	&$\left[10^{20.3},10^{22.0}\right)$	&$2.00$	&$24$	&$\left[10,100\right)$
\enddata
\tablenotetext{a}{The redshift evolution is parameterized by the effective optical depth $\tau_\mathrm{eff,\alpha}=B\left(1+z\right)^{\gamma+1}$ or the line density $\mathrm{d}n/\mathrm{d}z=A\left(1+z\right)^{\gamma}$.}
\tablenotetext{b}{The CDDF is a piecewise continuous power law $\mathrm{d}n/\mathrm{d}N_\mathrm{H\,I}\propto N_\mathrm{H\,I}^{-\beta}$.}
\tablenotetext{c}{The $b$ value distribution is $\mathrm{d}n/\mathrm{d}b\propto b^{-5}\exp\left(-b^4/b_\sigma^4\right)$.}
\end{deluxetable}

With our final set of input parameters (Table~\ref{simpar}) we computed
4000 MC line lists over the relevant redshift range $0\le z\le 4.6$. The number
of sightlines is large enough to reach convergence in the incidence of
optically thick \ion{H}{1} absorbers even at low redshifts
(Figs.~\ref{simsllsdla} \& \ref{dndz_logNge17p2}), thus providing sufficient
statistics for the highly stochastic UV LyC absorption.

\subsection{Mock quasar photometry}
\label{model_phot}

We used another Monte Carlo routine to generate mock quasar catalogs, i.e.\
distributions in emission redshift and observed magnitude, from the observed
luminosity function of quasars. Due to the strong attenuation by the IGM,
only quasars that are intrinsically bright in the continuum redward of
\ion{H}{1} Ly$\alpha$ can be detected with current UV instruments. Thus, we
adopted the SDSS DR3 luminosity function \citep{richards06} that is well
determined at bright magnitudes. We integrated their $z_\mathrm{em}>2.4$ pure
luminosity evolution model of the differential luminosity function in the
observed $i$ band at redshift two
$\phi\left(M_{i}^{z_\mathrm{em}=2},z_\mathrm{em}\right)$ combined with the
comoving volume in their adopted cosmological model to determine the all-sky
surface counts of quasars in a given range of redshift and absolute magnitude
\begin{equation}
\label{eq:qsocounts}
C_{4\pi}=\int_{z_\mathrm{min}}^{z_\mathrm{max}}\int_{M_{i,\mathrm{min}}^{z_\mathrm{em}=2}}^{M_{i,\mathrm{max}}^{z_\mathrm{em}=2}}\phi\left(M_{i}^{z_\mathrm{em}=2},z_\mathrm{em}\right)\frac{\mathrm{d}V(z_\mathrm{em})}{\mathrm{d}z_\mathrm{em}}\,\mathrm{d}M_{i}^{z_\mathrm{em}=2}\,\mathrm{d}z_\mathrm{em}.
\end{equation}
We chose to convert from absolute magnitude $M_{i}^{z_\mathrm{em}=2}$ to
$m_{1450}$, the observed AB magnitude at $1450$\AA\ in the quasar rest frame,
via the relation $M_{i}^{z_\mathrm{em}=2}=M_{1450}-1.486$ \citep{richards06},
yielding
\begin{equation}
m_{1450}=M_{i}^{z_\mathrm{em}=2}+5\log{\left(\frac{d_L}{\mathrm{Mpc}}\right)}-2.5\log{\left(1+z_\mathrm{em}\right)}+26.486
\end{equation}
with the luminosity distance $d_L(z_\mathrm{em})$. By varying the integration
limits of Equation~\ref{eq:qsocounts} we obtained a parameterization for
$\partial C_{4\pi}/\partial z_\mathrm{em}$ and
$\partial C_{4\pi}/\partial m_\mathrm{1450}$ which we used to simulate
$\sim 200000$ pairs $(z_\mathrm{em},m_\mathrm{1450})$ at $2.6<z_\mathrm{em}<4.6$
and $15<m_\mathrm{1450}<19$. This large mock sample ensured an accurate sampling
of the rare UV-bright population of quasars transparent at the \ion{He}{2} edge
(see \S\ref{results_counts} below). For comparison, Equation~\ref{eq:qsocounts}
predicts just $\sim 11000$ quasars on the full sky over the same range in
redshift and magnitude.

For each simulated quasar we assumed a unique spectral energy distribution
modeled as a power law $f_\nu\propto\nu^{-\alpha}$ with a break at \ion{H}{1}
Ly$\alpha$ \citep{telfer02}, normalized to yield the modeled $m_\mathrm{1450}$.
Redward of the break we assumed a Gaussian distribution of spectral slopes with
$(\left<\alpha_\mathrm{cont}\right>,\sigma\left(\alpha_\mathrm{cont}\right))=(0.5,0.3)$
whereas blueward of the break we assumed
$(\left<\alpha_\mathrm{UV}\right>,\sigma\left(\alpha_\mathrm{UV}\right))=(1.6,0.6)$
consistent with the large variation in far-UV spectral slopes found by
\citet{telfer02}\footnote{Note that \citet{telfer02} quote the standard error of
their mean spectral index instead of the (larger) standard deviation of the
distribution of spectral indices (their Fig.~11).}. To the quasar continua we
added the major quasar emission lines in the spectral range of interest
(Ly$\beta$, Ly$\alpha$, \ion{N}{5}, \ion{Si}{4}+\ion{O}{4}], \ion{C}{4}, \ion{C}{3}], \ion{Mg}{2}).
The emission lines were modeled as Gaussian profiles, the strengths and widths of
which were chosen consistent with \citet{vandenberk01}, with small variations
from quasar to quasar.

Lastly we blanketed each spectrum blueward of \ion{H}{1} Ly$\alpha$ by
\ion{H}{1} absorption in the IGM. For a given model quasar at a redshift
$z_\mathrm{em}$ we randomly drew one of our 4000 MC sightlines and computed the
\ion{H}{1} Lyman series and continuum absorption at $0<z<z_\mathrm{em}$
(\S~\ref{model_gen}), yielding a final mock quasar spectrum at
912\AA$<\lambda<$12000\AA. Blueward of \ion{He}{2} Ly$\alpha$ we assumed a
\ion{He}{2} Gunn-Peterson trough, resulting in zero flux (a reasonable
assumption since the GALEX FUV band covers the \ion{He}{2} break at
$z_\mathrm{em}>3.44$). We then obtained mock SDSS $ugriz$ photometry
\citep[asinh magnitudes,][]{lupton99} and mock GALEX FUV \& NUV photometry
(AB magnitudes) using the published filter curves \citep{morrissey05}. As
Galactic extinction becomes important in the UV, we also computed the
magnitudes after reddening each spectrum by the Galactic extinction curve
\citep{cardelli89}, adopting $R_V=3.1$ and a lognormal distribution in
$E(B-V)$ that closely resembles the color excess distribution towards SDSS
quasars \citep{schneider07}. At the high Galactic latitudes considered here,
the average extinction is $\simeq 0.26$\,mag and $\simeq 0.22$\,mag in the
FUV and NUV, respectively.

\section{Results}
\label{results}

\subsection{The expected number of UV-bright quasars at $z>2.7$}
\label{results_counts}

For each of our $\sim 200000$ model quasars we calculated the total \ion{H}{1}
Lyman continuum optical depth $\tau_\mathrm{LyC}$ at \ion{He}{2} Ly$\alpha$ in
the quasar rest frame, i.e.\ the accumulated \ion{H}{1} attenuation by
$\left[0.33\left(1+z_\mathrm{em}\right)-1\right]<z<z_\mathrm{em}$ absorbers.
This quantity characterizes the transparency of a sightline to the onset of
the \ion{He}{2} absorption, irrespective of lower-redshift LLSs that might
truncate the spectrum in the \ion{He}{2} forest region. We also computed the
AB magnitude of the quasar at \ion{He}{2} Ly$\alpha$
\begin{equation}
m_{304}=m_{1450}+0.191\alpha_\mathrm{cont}+1.506\alpha_\mathrm{UV}+1.086\tau_\mathrm{LyC},
\end{equation}
which depends on the input quasar magnitude $m_{1450}$, the spectral slopes of
the continuum blueward ($\alpha_\mathrm{UV}$) and redward
($\alpha_\mathrm{cont}$) of \ion{H}{1} Ly$\alpha$ and $\tau_\mathrm{LyC}$.

\begin{figure}
\includegraphics[width=\columnwidth]{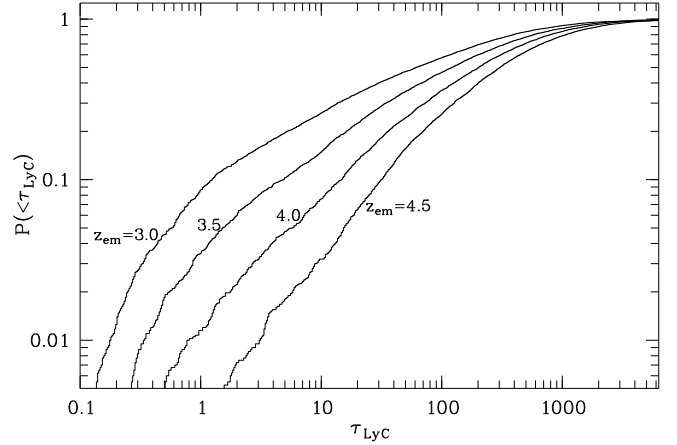}
\caption{\label{taulycstat}Predicted cumulative probability distributions of the
\ion{H}{1} Lyman continuum optical depth $\tau_\mathrm{LyC}$ at \ion{He}{2}
Ly$\alpha$ in the rest frame of a source at redshift $z_\mathrm{em}$.}
\end{figure}

In Fig.~\ref{taulycstat} we plot the cumulative distribution function of
$\tau_\mathrm{LyC}$ from our 4000 MC sightlines for different quasar emission
redshifts. Our calculations indicate a very low probability to encounter a
sightline that is not highly attenuated at the \ion{He}{2} edge, consistent with
previous estimates \citep{picard93,jakobsen98}. The accumulated continuum optical
depth strongly increases with emission redshift. While $\simeq 9$\% of all
quasars at $z_\mathrm{em}=3$ should be 'transparent' ($\tau_\mathrm{LyC}<1$),
this fraction drops to $\sim 1$\% at $z_\mathrm{em}=4$.

\begin{figure}
\includegraphics[width=\columnwidth]{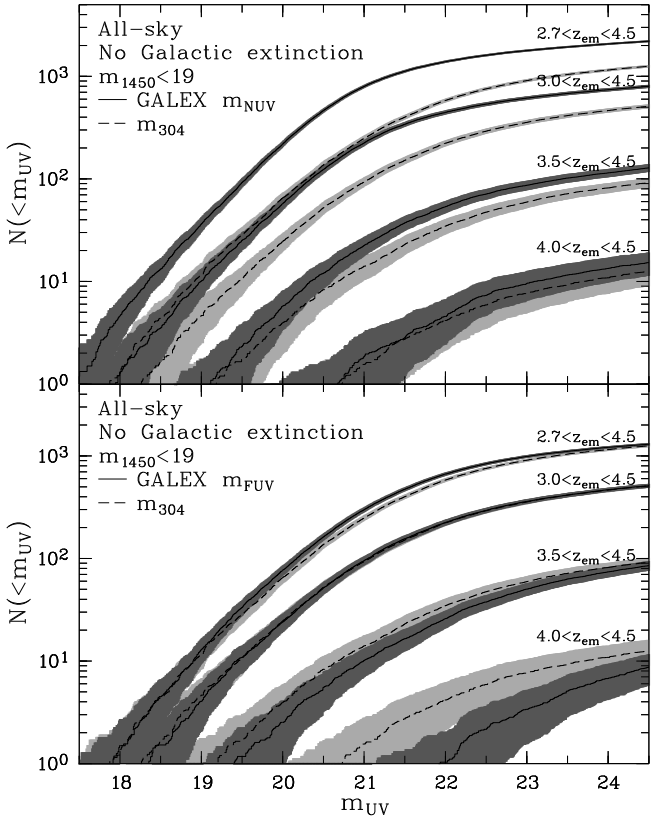}
\caption{\label{allskycounts}Predicted cumulative all-sky number counts of
$m_{1450}<19$ quasars as a function of limiting UV magnitude for various
redshift ranges. The upper (lower) panel compares the predicted source counts in
the GALEX NUV (GALEX FUV) band (solid lines) to the ones inferred at \ion{He}{2}
Ly$\alpha$ in the source rest frame (shown dashed in both panels), respectively.
The gray shaded regions indicate $1\sigma$ errors in the source counts due to
cosmic variance. Note that these predictions include the attenuation by the IGM,
but they have not been corrected for Galactic extinction.}
\end{figure}

\begin{figure*}
\includegraphics[width=\textwidth]{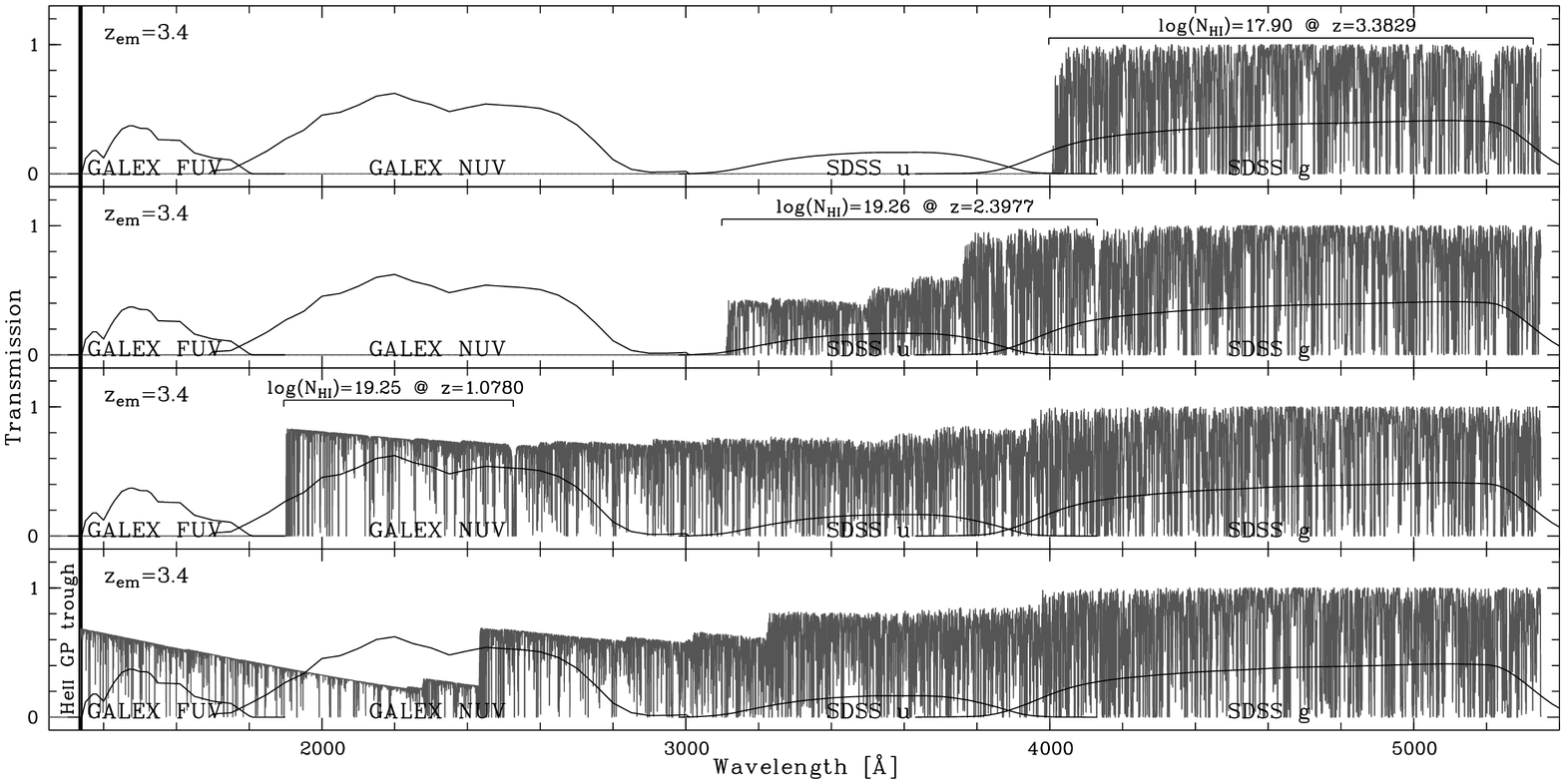}
\caption{\label{examplelos}Four simulated normalized \ion{H}{1} forest Lyman
series and Lyman continuum absorption spectra of a source at $z_\mathrm{em}=3.4$
with overplotted filter bandpasses. The upper three panels show sightlines with
optically thick systems that result in photometric dropouts without recovery at
the \ion{He}{2} Ly$\alpha$ edge (vertical line). In the sightline shown in the
uppermost panel, an intervening SLLS prevents the spectrum from recovering from
the first encountered LLS break. The sightline in the lowest panel does not have
an intervening optically thick absorber, so that the background source is
detectable in all photometric bands until the onset of the assumed \ion{He}{2}
Gunn-Peterson trough.}
\end{figure*}

The predicted number of high-$z$ quasars detectable in the far UV primarily
depends on the increasing opacity and the declining quasar space density at
$z_\mathrm{em}\ga 3$. Figure~\ref{allskycounts} shows the predicted cumulative
all-sky number counts of $m_{1450}<19$ quasars in the GALEX FUV \& NUV bands
compared to their predicted $m_{304}$ for various redshift ranges. These
estimates have not been corrected for Galactic extinction, in particular close
to the Galactic plane. For an $E(B-V)>1$ commonly encountered at Galactic
latitudes $|b|\la 20\degr$, the FUV extinction is $>8$\,mag, so that $\sim 25$\%
of the sky are effectively blocked for \ion{He}{2} studies even if quasars are
found in this 'Zone of Avoidance' \citep{hubble34}.

The SDSS luminosity function predicts $\sim 9200$ $m_{1450}<19$ quasars on the
entire sky at $2.7<z_\mathrm{em}<4.5$ (eq.~\ref{eq:qsocounts}). More than 200
of these should have $m_{304}<21$, well within the capabilities of HST. However,
at $4<z_\mathrm{em}<4.5$ there should be just $\sim 600$ $m_{1450}<19$ quasars
on the whole sky, the sightlines of which encounter larger LyC attenuation,
yielding just $\sim 1$ quasar at $m_{304}<21$. At these high redshifts cosmic
variance has a strong impact on the real number counts. The same is true for the
least-attenuated UV-brightest quasars that are located at the lowest redshifts.
In order to obtain accurate results both at the highest redshifts and the
brightest UV magnitudes we had to simulate the large set of $\sim 200000$ quasars,
corresponding to $\sim 20\times$ the predicted all-sky number counts.

The GALEX bands trace the small UV-transparent quasar population very well,
but differently at different redshifts. At $z_\mathrm{em}<3.5$ the GALEX NUV
band is not a good indicator for the flux at \ion{He}{2} Ly$\alpha$ because of
the high probability to encounter a Lyman limit break in the large wavelength
range between the NUV band and the onset of \ion{He}{2} absorption. As the
FUV band is closer to the \ion{He}{2} edge it is a more sensitive indicator of
flux at \ion{He}{2} Ly$\alpha$. At $z_\mathrm{em}>3.44$ the FUV band samples
the \ion{He}{2} edge and the presumed \ion{He}{2} Gunn-Peterson trough. The
\ion{He}{2} Ly$\alpha$ absorption progressively attenuates the FUV flux and
likely causes FUV dropouts at $z_\mathrm{em}>4$. Only at $z_\mathrm{em}>4$
will the GALEX NUV flux indicate a likely transparent sightline.

Figure~\ref{examplelos} further illustrates the importance of detected FUV flux
to select promising sightlines for \ion{He}{2} absorption. We show the normalized
\ion{H}{1} Lyman series and Lyman continuum transmission spectra of four
representative mock sightlines from the onset of the \ion{He}{2} Gunn-Peterson
trough to \ion{H}{1} Ly$\alpha$ at the emission redshift $z_\mathrm{em}=3.4$.
In the sightlines shown in the upper three panels the indicated optically thick
\ion{H}{1} absorbers truncate the spectra at the Lyman limit, causing dropouts
in the overplotted filter bands. Obviously, only $z_\mathrm{em}\sim 3.4$ quasars
detected in the GALEX FUV band will show a transparent sightline that has
recovered from intervening LLS breaks. Even for the small subset of high-$z$
quasars detected by GALEX, intervening low-redshift LLSs likely truncate the
quasar flux between the two GALEX bands. Thus, in order to select transparent
sightlines at a high success rate, FUV detections are required at least at
$z_\mathrm{em}<3.4$ where the FUV band still samples the quasar continuum
redward of \ion{He}{2} Ly$\alpha$.

\subsection{Far-UV color selection of probable \ion{He}{2} sightlines}
\label{sect_finalsample}

\begin{figure*}\centering
\includegraphics[width=0.9\textwidth]{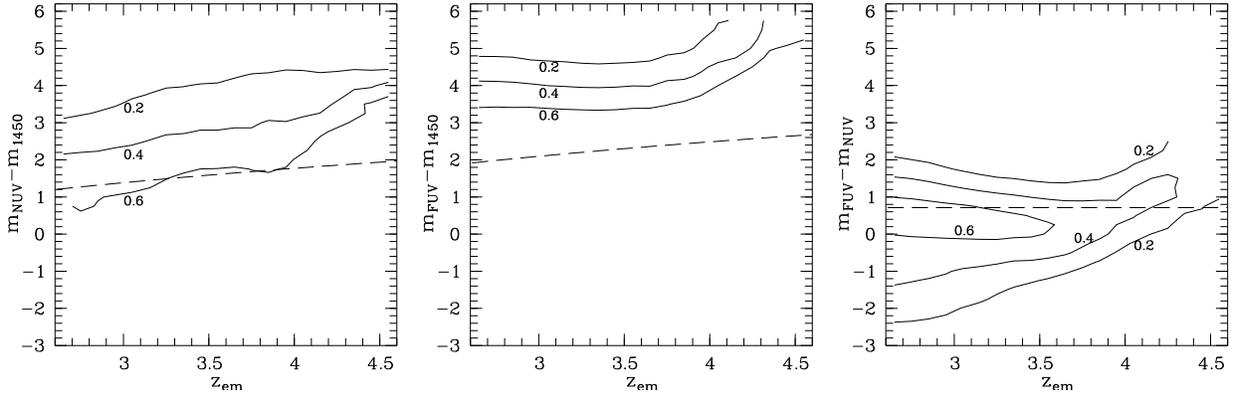}
\caption{\label{galexselmap}Predicted colors of quasars detectable with GALEX as
a function of emission redshift $z_\mathrm{em}$ (\emph{left:}
$m_\mathrm{NUV}-m_\mathrm{1450}$; \emph{middle:} $m_\mathrm{FUV}-m_\mathrm{1450}$;
\emph{right:} $m_\mathrm{FUV}-m_\mathrm{NUV}$). The contours delineate the
probability that a quasar at a given redshift $z_\mathrm{em}$ detected by GALEX
at a given color will be transparent at the \ion{He}{2} 304\AA\ edge
($\tau_\mathrm{LyC}<1$). The dashed lines mark the colors of the mean adopted
quasar spectral energy distribution (\S\ref{model_phot}) ignoring intergalactic
absorption.}
\end{figure*}

Figure~\ref{examplelos} also illustrates that the GALEX UV color
$m_\mathrm{FUV}-m_\mathrm{NUV}$ can be used to select the most promising
sightlines to discover \ion{He}{2} absorption. Significantly red GALEX colors
indicate low-$z$ LLS breaks (3rd panel of Fig.~\ref{examplelos}) between the
FUV and the NUV band, whereas blue GALEX colors signal the recovery from a LLS
break or the relatively unabsorbed hard quasar continuum. NUV-only detections
indicate transparent sightlines only if the FUV band significantly covers the
strong \ion{He}{2} absorption, i.e.\ at very high redshift
($z_\mathrm{em}\ga 4$).

We used our mock quasar photometry to determine the fraction of transparent
sightlines (defined as the fraction of sightlines with $\tau_\mathrm{LyC}<1$)
as a function of redshift. In Figure~\ref{galexselmap} we plot the probability
contours that a quasar detected by GALEX at a given color will show a total
$\tau_\mathrm{LyC}<1$ along the line of sight at \ion{He}{2} Ly$\alpha$ in the
quasar rest frame. The UV-optical colors $m_\mathrm{NUV}-m_{1450}$ and
$m_\mathrm{FUV}-m_{1450}$ just give modest hints whether the quasar will show
flux at \ion{He}{2} Ly$\alpha$. The NUV-optical color indicates a transparent
sightline just at the highest redshifts, but is otherwise quite insensitive
due to the frequent low-$z$ LLS breaks between the NUV band and the \ion{He}{2}
edge. Thus, at any redshift the least-absorbed quasars with the bluest
$m_\mathrm{NUV}-m_{1450}$ colors are the most promising candidates to detect
\ion{He}{2}. The FUV-optical color $m_\mathrm{FUV}-m_{1450}$ provides better
constraints. Quasars at $z_\mathrm{em}<3.4$ at a $m_\mathrm{FUV}-m_{1450}\la 3.4$
have a $\ga 60$\% chance to show a low $\tau_\mathrm{LyC}<1$ along the line of
sight. At higher redshifts, the \ion{He}{2} Gunn-Peterson trough reddens the
FUV-optical color.

The UV color $m_\mathrm{FUV}-m_\mathrm{NUV}$ (right panel of
Fig.~\ref{galexselmap}) yields the most natural color-selection constraints. Any
quasar detected in both GALEX bands at a rather blue UV color has a high chance
to show flux at \ion{He}{2} Ly$\alpha$. Unless the FUV fluxes get severely
absorbed by \ion{He}{2} at $z_\mathrm{em}\ga 4$, the GALEX UV colors of
transparent quasars should be similar to those of their unabsorbed spectral
energy distributions, with the slightly bluer colors indicating the recovery
from partial LLSs that result in a steeply rising flux towards the FUV due to
the strong frequency dependence of the LyC cross-section. Quasars at
$m_\mathrm{FUV}-m_\mathrm{NUV}\ga 2$ are likely to show a LLS break at the blue
end of the FUV band even if they are detected in the FUV. Very blue quasars
below the lower 20\% line in the right panel of Fig.~\ref{galexselmap} are
recovering from a $\tau_\mathrm{LL}>1$ LLS break at $z>2$ so that their flux
rises steeply in both GALEX bands.

\begin{figure*}
\includegraphics[width=\textwidth]{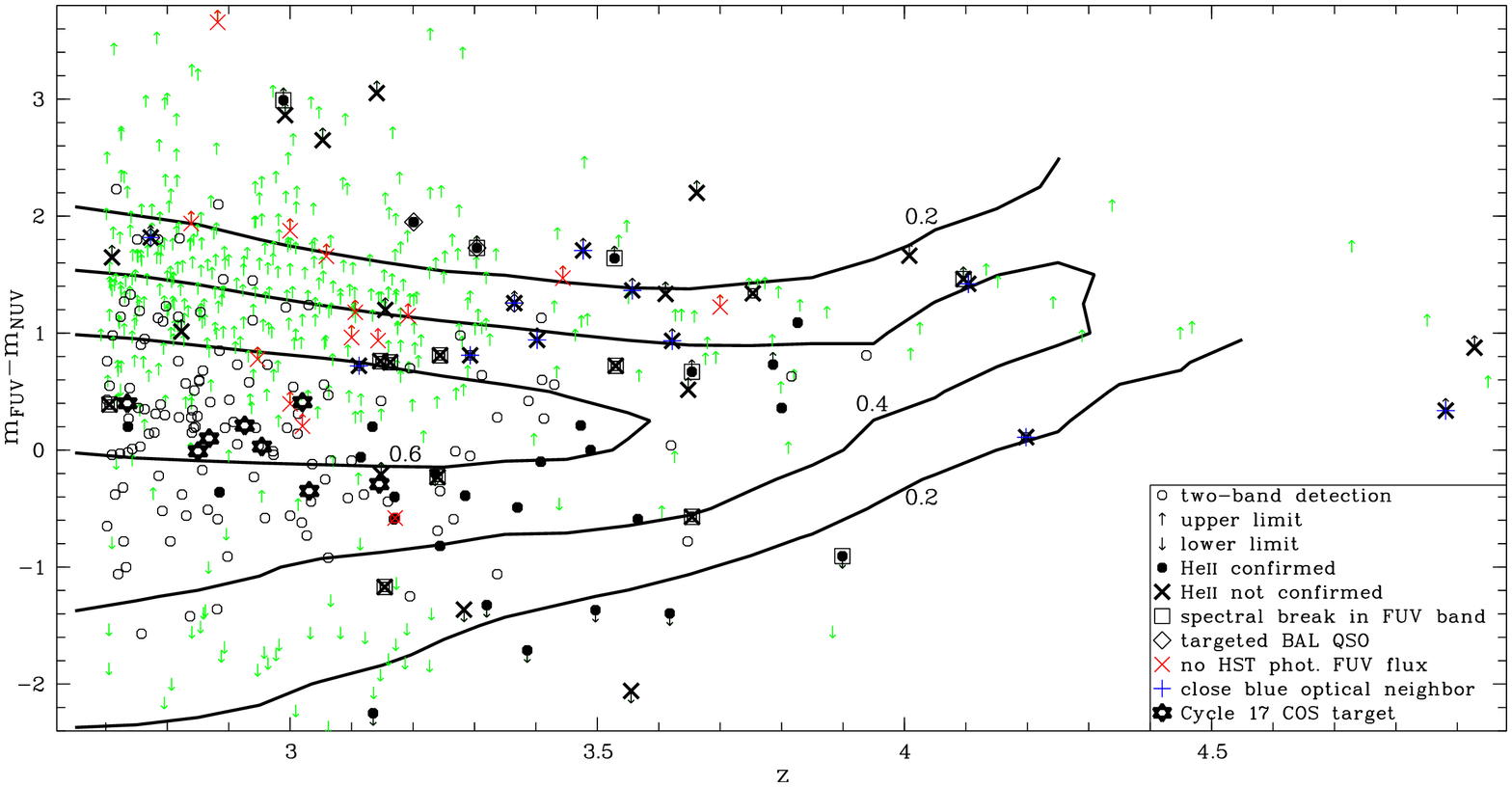}
\caption{\label{gr4zcolor}GALEX UV color $m_\mathrm{FUV}-m_\mathrm{NUV}$ of
$z_\mathrm{em}>2.7$ quasars in the GALEX GR4 source catalog. For quasars in the
SDSS footprint without UV follow-up observations we only show those that do not
have sufficiently blue neighboring objects out to 5\arcsec\ separation to avoid
GALEX source confusion. Open circles show quasars detected in both GALEX bands,
whereas arrows indicate upper (lower) color limits for sources detected in the
NUV (FUV) only. Filled circles indicate the UV colors of quasar sightlines
spectroscopically confirmed to be transparent at \ion{He}{2} Ly$\alpha$.
Quasars showing low spectroscopic FUV flux, either due to a Lyman limit break
in the FUV or a recovery from a Lyman limit break in the NUV, are shown as open
squares. Thick crosses mark quasars targeted with HST, but not confirmed to
show flux at \ion{He}{2} Ly$\alpha$, either because of optically thick Lyman
limit breaks in the FUV spectral range (crossed squares) or no detectable FUV
flux at all (thick $\times$ crosses). Quasars without significant FUV flux in
follow-up HST images are indicated as well (thin $\times$ crosses). Some
spectroscopically observed quasars exhibit BAL features (diamonds), and several
previous follow-up observations seem to have been affected by GALEX source
confusion ($+$ signs). Thick star symbols mark the 8 UV-bright quasars we
selected for upcoming follow-up spectroscopy with HST/COS in Cycle 17. The
thick lines show the probability contours that a quasar at a given UV color
shows flux down to the onset of the \ion{He}{2} absorption, based on our MC
simulations.}
\end{figure*}

With these estimates on the UV color range of quasars that show flux at
\ion{He}{2} Ly$\alpha$, we can estimate \ion{He}{2} detection probabilities for
the actual GALEX-detected $z_\mathrm{em}>2.7$ quasars. Figure~\ref{gr4zcolor}
compares the GALEX $m_\mathrm{FUV}-m_\mathrm{NUV}$ colors of our transparent
($\tau_\mathrm{LyC}<1$) mock quasars to actual observations. We find that the UV
colors of quasars having sightlines that are known to be transparent down to the
onset of \ion{He}{2} absorption are similar to the simulated UV colors of
transparent quasars. A posteriori, the blue UV colors of most known \ion{He}{2}
quasars indicate a high probability for transparency. Among the
GALEX-detected quasars without further follow-up the rare quasars at
$m_\mathrm{FUV}-m_\mathrm{NUV}\la 1$ are the best candidates to search for flux
at \ion{He}{2} Ly$\alpha$. Our MC simulations indicate a probability of
$\ga 60$\% that a $z_\mathrm{em}\la 3.5$ quasar detected at
$0\la m_\mathrm{FUV}-m_\mathrm{NUV}\la 1$ will show $\tau_\mathrm{LyC}<1$.
The slight offset between the simulated and the observed UV colors of transparent
quasars could be due to a generally harder UV spectral energy distribution than
assumed in the simulations (i.e.\ $\left<\alpha_\mathrm{UV}\right><1.6$) and/or
a higher mean LyC absorption from a larger population of $z\la 2$ LLSs.
We suspect the latter is more likely given the poor existing constraints on the
exact CDDF and the evolution of the MFP at $z<3.6$ (\S\ref{model_cddf}).

In contrast, quasars confirmed by HST follow-up to show zero flux at \ion{He}{2}
Ly$\alpha$ are mostly redder in $m_\mathrm{FUV}-m_\mathrm{NUV}$ than the
UV-transparent population, consistent with our simulations. Especially the high
upper limits $m_\mathrm{FUV}-m_\mathrm{NUV}\ga 2$ correspond to significant
detections in the NUV, but no formal detection in the FUV, signaling the cutoff
by an optically thick LLS. The only opaque sightlines that remain insensitive to
our UV color selection are the ones intercepted by a LLS just within the narrow
range between the blue end of the FUV bandpass and \ion{He}{2} Ly$\alpha$
(e.g.\ PKS~1442$+$101 in Fig.~\ref{hstspc}). This is reflected in our simulations
by the broadening color contours towards lower redshifts.

If the LLS is not optically thick then the flux can recover, but the quasar is
of very limited scientific value because it is too faint for follow-up at
\ion{He}{2} Ly$\alpha$ (i.e.\ $\tau_\mathrm{LyC}>1$; \citealt{syphers09b,syphers09a}
have identified two such quasars). Moreover, the two BALQSOs confirmed in the
FUV by \citeauthor{syphers09a} show red GALEX UV colors, presumably due to their
intrinsically redder spectral energy distributions and/or BAL troughs extending
in the UV. While these quasars are interesting to study the BAL phenomenon, they
are effectively useless for investigating intergalactic \ion{He}{2} absorption as
one cannot distinguish IGM \ion{He}{2} Gunn-Peterson troughs from potential BAL troughs. 

The UV color separates well between blue \ion{He}{2}-transparent quasars and red
opaque ones, despite the low S/N near the GALEX detection limit. However, quasars
just detected in one of the GALEX bands require further attention. FUV-only detected
sightlines probably recover from a partial LLS break so that the low NUV flux is
beyond the detection limit. Given that we just quote $1\sigma$ flux limits on NUV
dropouts, the colors of the 6 very blue confirmed \ion{He}{2} quasars could be
similar to those of the other \ion{He}{2} quasars. Likewise, the FUV flux of some
transparent quasars detected just in the NUV should have been detected as well.
Nevertheless, since significant NUV-only detections indicate opaque sightlines,
such background quasars should not be regarded as prime candidates for spectroscopic
follow-up. Generally, we do not consider very low S/N$<2$ detections in a single
GALEX band to be real, whereas sources detected in both GALEX bands probably are,
as the GALEX pipeline performs the source detection independently before merging the
catalogs \citep{morrissey07}.

Moreover, quasars with nearby optical neighbors should be avoided, as they will
be likely affected by GALEX source confusion due to the broad instrument PSF.
Apart from GALEX-detected $z_\mathrm{em}>2.7$ quasars with HST follow-up,
Fig.~\ref{gr4zcolor} shows only those sources which qualify for further
investigation (non-BAL, no blue neighboring source in SDSS DR7 at separation
$<5\arcsec$). Given that the UV color is not well constrained at low S/N
(a S/N$>3$ in both bands corresponds to $\sigma(m_\mathrm{FUV}-m_\mathrm{NUV})<0.51$)
we consider two subsets of these quasars as the most promising ones for further
detections of \ion{He}{2}: (i) those 52 which have been significantly detected
in both bands at S/N$>3$ and have $m_\mathrm{FUV}-m_\mathrm{NUV}<1$, and (ii)
the 114 remaining quasars detected at S/N$>2$ in the FUV band. These samples
are presented in Tables~\ref{candlist1} and \ref{candlist2}, respectively.
We caution that several GALEX detections outside the SDSS DR7 footprint will
correspond to confused GALEX sources ($\sim 20\%$ if we adopt our estimate from SDSS).
Likewise, a few quasars with red optical neighbors (flagged in Tables~\ref{candlist1}
and \ref{candlist2}) might be confused sources, since our neighbor classification
was based on the broadband SED shape from SDSS and GALEX.
Seven quasars detected on DIS survey plates have been flagged as potentially affected
by source confusion. The majority (90\%) of the $z_\mathrm{em}>2.78$ quasars in
Tables~\ref{candlist1} and \ref{candlist2} were previously suggested as candidate
\ion{He}{2} quasars by \citet{syphers09b}. All but one of the 13 additional
$z_\mathrm{em}>2.78$ quasars have GALEX counterparts beyond the match radius adopted
by \citealt{syphers09b} (3\arcsec).

\section{Application to the Sloan Digital Sky Survey}
\label{sdssgalex}

\subsection{Comparing UV-bright SDSS quasars to predictions}
With a homogeneous well-characterized large area quasar survey such as SDSS, we
can compare our predicted number counts of UV-bright quasars to actual
observations after accounting for several observational effects. First, the
predicted all-sky number counts (Fig.~\ref{allskycounts}) were corrected for
Galactic foreground extinction as incorporated in our calculations
(\S\ref{model_phot}). Then we accounted for the actual GALEX GR4 sky coverage
and depth. The exposure time varies significantly among tiles of a given GALEX
imaging survey, rendering them inherently inhomogeneous. Therefore we used the
GALEX instrument sensitivity \citep{morrissey07} and the actual GR4 tile
exposure times to calculate a $5\sigma$ limiting magnitude for each tile. With
an approximate area correction for the overlapping circular GALEX tiles
\citep[e.g.][]{budavari09} we then calculated the GR4 sky coverage as a function
of limiting magnitude. We regard the S/N$\ge 5$ threshold as sufficient to avoid
incompleteness in the GALEX source catalog, but we note that apart from general
source counts \citep{bianchi07} the repeatability and S/N stability of GALEX is
not well established at its instrumental limit. 

Next, we accounted for the SDSS sky coverage. Considering that GALEX GR4 covers
almost the full sky at high Galactic latitude ($|b|\ga 20\degr$), we avoided the
cumbersome calculation of the actual overlapping area of SDSS DR7 and GALEX GR4
(see \citealt{budavari09} for an application to DR6+GR3), and adopted instead
the SDSS Legacy spectroscopic sky coverage of 8032\,deg$^2$. SEGUE fields were
not taken into account, as they are mainly at low Galactic latitude and have a
significantly smaller quasar targeting rate. Lastly, we corrected for the SDSS
quasar selection efficiency to predict the number of UV-bright SDSS quasars
detectable with GALEX. As SDSS selects quasars primarily by color, we used the
photometric SDSS selection function by \citet{richards06} averaged at $i<19.1$.
The magnitude cut provides a homogeneous survey limit at $z_\mathrm{em}>2.7$
(SDSS selects $z\ga 3$ quasar candidates at $i<20.2$) and ensures that the
selection function does not depend on magnitude. Moreover, it is well matched
to the rest-frame magnitude limit we applied in our simulations
($i\sim m_{1450}<19$), as the $i$ band covers the quasar continuum redward of
Ly$\alpha$ at the relevant redshifts. The different bandpasses induce a slight
redshift-dependent offset $i-m_{1450}\sim -0.1$, but uncertainties in the $K$
correction used to determine the quasar luminosity function are larger than this.

\begin{figure}
\includegraphics[width=\columnwidth]{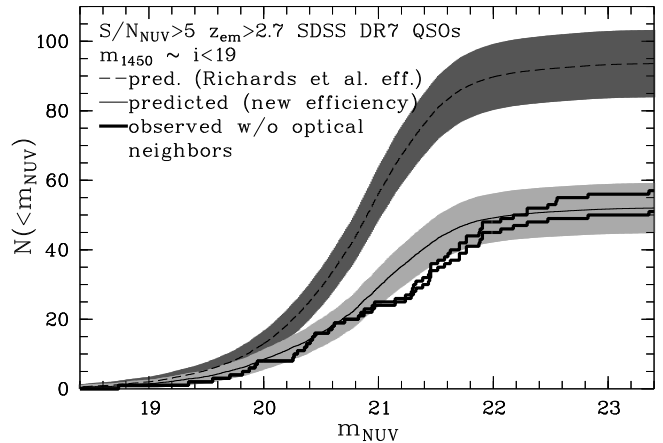}
\caption{\label{comparecounts}Comparison of predicted and observed cumulative
number counts of $z_\mathrm{em}>2.7$ $i\la 19$ SDSS DR7 quasars detectable with
GALEX at S/N$>5$ in the NUV. The thin lines plot the predicted number counts
using the SDSS color selection efficiency from \citet[][dashed]{richards06},
and our color-dependent selection efficiency (solid), respectively. Shaded
regions outline Poisson errors on the source counts. The thick solid lines show
the observed cumulative number counts of SDSS DR7 w/o potentially confused GALEX
detections due to nearby blue optical neighbors.}
\end{figure}

From our sample of quasars we then selected only those 58 which were
targeted by SDSS, have $i<19.0$ and have been detected by GALEX at S/N$>5$ in
the NUV band. If we exclude SDSS quasars with blue optical neighbors that could
be cases of GALEX source confusion, this number reduces to 52.
Figure~\ref{comparecounts} compares the cumulative number counts of these
NUV-detected quasars to the prediction based on our IGM model and the SDSS
selection function by \citet[][\emph{upper} curve]{richards06}. Adopting the
\citet{richards06} selection efficiency, the number of NUV-detectable quasars is
a factor of $\sim 2$ larger than observed, even including potentially confused
GALEX sources. The predicted number counts can only be lowered by increasing the
LyC opacity in our IGM model or by decreasing the SDSS selection efficiency.
The uncertainties on other model ingredients, such as the luminosity function of
bright quasars, the $K$ correction, and the GALEX+SDSS footprint corrections,
are too small to create this discrepancy. Given that our IGM model fits the MFP
measurements (Fig.~\ref{mfpmatch}) and independently reproduces the observed
redshift evolution of LLSs (Fig.~\ref{dndz_logNge17p2}) we have focused here on
systematic effects in the SDSS selection efficiency.

\subsection{A color-dependent SDSS color selection function}

Quasar selection by broadband colors is expected to be inefficient and highly
model-dependent at $z\sim 3$, where quasar colors are similar to those of
main-sequence stars \citep[e.g][]{richards06}. We used the simulated SDSS
photometry of our $\sim 200000$ $z_\mathrm{em}>2.6$ model quasars
(\S\ref{model_phot}) to reassess the SDSS quasar selection function. SDSS
selects most quasar candidates as outliers from the stellar locus in
multi-dimensional color space. Because this procedure depends on the photometric
errors, we computed these by fitting the photometric errors of observed
$z_\mathrm{em}>2.7$ SDSS DR5 quasars \citep{schneider07} as a function of
magnitude. We associated each mock SDSS magnitude $m$ with the fitted mean
photometric error $\sigma_m$ without modifying the mock magnitude. Thus, we
assume perfect SDSS photometry with a realistic mean error, which simplifies
our further discussion, but will likely result in an overestimate of the
selection efficiency due to photometric uncertainties near the SDSS survey limit,
particularly in the $u$ band. Potential effects of asymmetric distribution
functions of SDSS magnitudes and their errors at the survey limit are beyond
the scope of this paper. 

Gordon Richards kindly agreed to process our mock photometry with the final SDSS
quasar target selection algorithm \citep{richards02b} that incorporates the
imposed 10\% follow-up targeting rate of quasars whose colors intersect the
stellar locus (the 'mid-$z$' inclusion box of \citealt{richards02b}). The result
of that operation is a selection flag for each mock quasar indicating whether it
would have been targeted under SDSS routine operations. We then computed average
SDSS selection efficiency as the fraction of selected mock quasars in
$\Delta z_\mathrm{em}=0.05$ bins.

\begin{figure}
\includegraphics[width=\columnwidth]{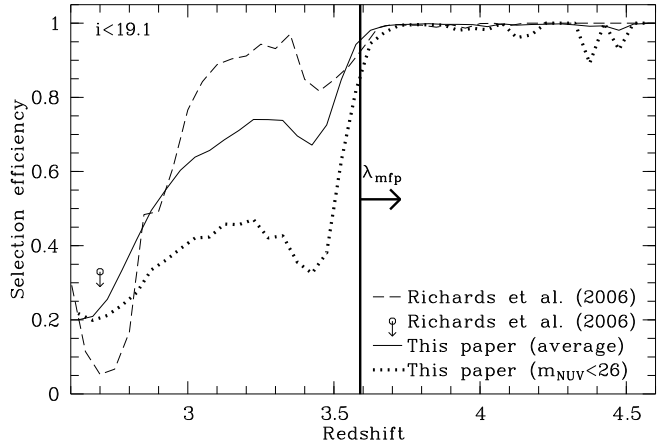}
\caption{\label{sdssselfunc}Average SDSS quasar color selection efficiency
derived using our MC simulations of SDSS photometry (solid line) compared to the
previous estimate by \citeauthor{richards06} (dashed line). The average
selection efficiency is almost magnitude-independent for the bright $i<19.1$
quasars considered here. The open circle marks the lower limit on the
$z\simeq 2.7$ selection efficiency estimated by \citet{richards06} by requiring
a smooth luminosity function at lower and higher redshifts. The thick dotted
line shows the SDSS color selection efficiency of $m_\mathrm{NUV}<26$ quasars.
The vertical bar marks the start of the first redshift bin we regarded as
unbiased for the measurement of the MFP \citep{prochaska09}.}
\end{figure}

In Fig.~\ref{sdssselfunc} we compare our selection function to the one by
\citet{richards06}. Both selection functions are essentially unity at
$z_\mathrm{em}\ga 3.6$ where colors of quasars are sufficiently red because of
IGM absorption to separate well from the stellar locus. In particular,
high-redshift LLSs will result in red $u-g$ colors due to $u$ band dropouts
(Fig.~\ref{examplelos}). At $z_\mathrm{em}\la 3.5$, however, there is a striking
difference between the two selection functions. Our average selection efficiency
at $z_\mathrm{em}\simeq 3.2$ is $\sim 25$\% smaller than the one by
\citet{richards06}, whereas at $z_\mathrm{em}\simeq 2.7$ it is a factor of
$\sim 4$ higher, and is in better agreement with their upper limit based on the
expected smoothness of the luminosity function with redshift. The main model
ingredients affecting the colors, and thus the selection efficiency, are the
quasar spectral energy distributions and the IGM, i.e.\ the LyC absorption.
Apart from a larger spread in the power-law spectral index blueward of
\ion{H}{1} Ly$\alpha$, our parameters to model the intrinsic quasar spectra are
very similar to the ones used by \citet{richards06}, so these discrepancies must
be due to different assumptions regarding the properties of the IGM that result
in statistically different quasar colors.

\begin{figure}
\includegraphics[width=\columnwidth]{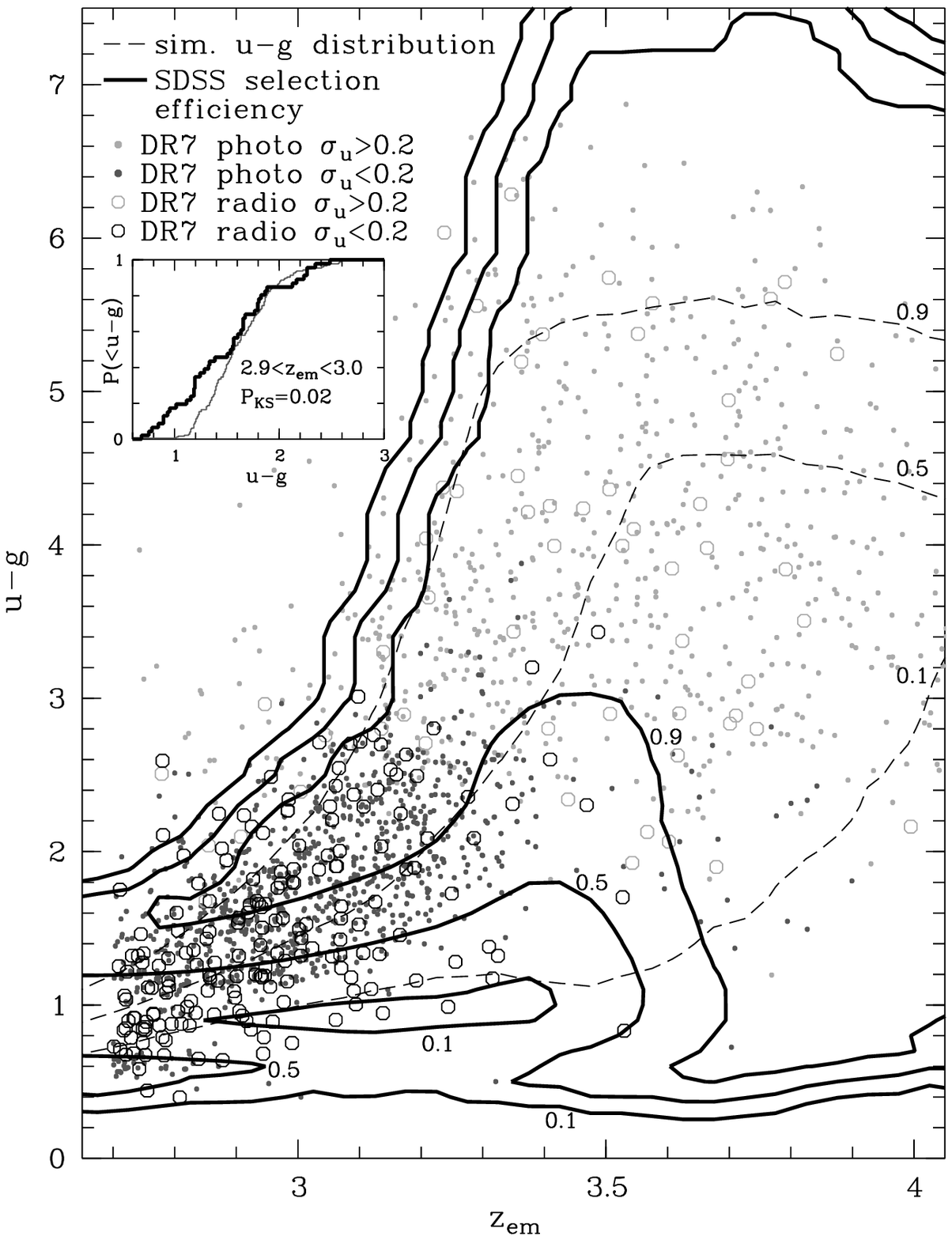}
\caption{\label{sdss_u_minus_g}Mock vs. observed $u-g$ quasar color as a
function of redshift. Small filled circles (large open circles) show
color-selected (radio-selected) $i<19.1$ quasars from the SDSS DR7 quasar catalog
\citep{schneider10}. All plotted quasars have been selected by the final SDSS
criteria \citep{richards02b}. Quasars marginally detected in the $u$ band
($\sigma_u=0.2$ corresponds to S/N$\simeq 5$) are plotted in light gray. The
dashed lines show the 10\%, 50\% and 90\% percentiles of the $u-g$ color
distribution of the simulated quasars. The solid contours show our derived SDSS
selection efficiency at a given redshift and $u-g$ color. At
$3.2\la z_\mathrm{em}\la 3.6$ blue ($u-g\la 2$) quasars are missed by the SDSS
color criteria, in contrast to radio-loud quasars selected independent of color.
The lines in the inset compare the cumulative probability distributions of
color-selected (thin) and radio-selected (thick) $2.9<z_\mathrm{em}<3$ SDSS quasars.
The latter exhibit systematically bluer $u-g$ color than the former.}
\end{figure}

The selection efficiency of our model quasars critically depends on the $u-g$
color. Figure~\ref{sdss_u_minus_g} compares the distribution of mock $u-g$
quasar colors to observed $i<19.1$ SDSS DR7 quasars \citep{schneider10}, either
selected based solely on the \citet{richards02b} color selection criteria, or on
their radio flux. SDSS targets radio-detected quasar candidates independently of
color. The color-selected quasars have significantly redder $u-g$ colors than
the radio-selected ones at all redshifts $z_\mathrm{em}>2.7$ allowing for such a
comparison (see the inset in Fig.~\ref{sdss_u_minus_g}). They are also redder
than most of our simulated quasars at $z_\mathrm{em}\la 3.5$, whereas the
radio-selected quasars fill the simulated range in $u-g$ color. We verified that
most quasars with very red $u-g$ colors outside the simulated range are BALQSOs
that were not treated by our MC simulations
(for the selection efficiency of BALQSOs see \citealt{allen10}).

The characteristic shape of the simulated color distribution is due to the SDSS
magnitude system \citep{lupton99} that yields finite values even for zero or
negative fluxes. At $z_\mathrm{em}>3.4$ the frequent LLSs result in $u$ band
dropouts with a finite $u=24.63$ at zero flux as defined for SDSS. At higher
redshifts the $g$ band flux is progressively attenuated by the IGM, which
results in an artificially blue $u-g$ color if the $u$ band flux is zero. The
$u-g$ colors of $z_\mathrm{em}>3.4$ quasars are not well determined as the $u$
magnitude exceeds usually employed detection limits. In this regime, the $u$
band flux is systematically overestimated due to Eddington bias, so that the
observed $u-g$ colors are bluer than the simulated ones without Eddington bias.
Considering that SDSS selects even fainter $i<20.2$ high-redshift candidates,
systematic effects in their colors at the faint end of the survey may
non-trivially alter the selection function. Such effects are best explored by
photometric analysis of simulated survey images \citep[e.g.][]{hunt04,glikman10}.

The thick contours in Fig.~\ref{sdss_u_minus_g} show the SDSS selection
efficiency at a given quasar redshift as a function of the $u-g$ color. At high
redshifts ($z_\mathrm{em}\ga 3.6$) the large range in color with a high
selection efficiency means that almost all simulated quasars are selected
regardless of their $u-g$ color. However, at $3\la z_\mathrm{em}\la 3.5$ the
SDSS quasar targeting algorithm preferably selects red quasars and
systematically misses blue ones. This \emph{color-dependent} selection
efficiency is in good agreement with the distribution of the observed
color-selected SDSS quasars in Fig.~\ref{sdss_u_minus_g}. In particular, very
few observed quasars have $u-g<1$ at $z_\mathrm{em}>3$, and most
$z_\mathrm{em}\simeq 3.4$ SDSS quasars have $u-g>2$, leaving a prominent 'hole'
in color space compared to our predictions. On the other hand, the
radio-selected SDSS quasars still reside in the color range of low selection
efficiency. Our simulations also recover inhomogeneities in the color selection
of $z_\mathrm{em}\la 3$ quasars. \citet{richards02b} define the 'mid-$z$'
inclusion box at $0.6<u-g<1.5$ with a targeting rate limited to 10\% due to
overlap with the stellar locus. However, candidates having $u-g<0.6$ are always
followed up (this is the UV-excess criterion of \citealt{richards02b}). Hence,
there is a 'cluster' of DR7 quasars at $u-g\simeq 0.6$ selected by UV excess,
whereas at $0.7<u-g\la 1$ there are very few color-selected quasars.

The color-dependent selection efficiency of SDSS is due to the difficulty to
differentiate quasar colors from stellar colors. The blue quasars at
$3\la z_\mathrm{em}\la 3.5$ do not separate well from the stellar locus, hence
they are preferentially missed by the SDSS color selection criteria. But how
does this explain the difference in the selection functions? \citet{richards06}
used the IGM model by \citet{fan99} that results in significantly redder $u-g$
colors and a high selection efficiency of $z_\mathrm{em}\ga 3$ quasars and,
therefore, in a higher predicted selection efficiency. An explicit color
dependence of the selection efficiency complicates 'completeness' corrections of
color-based quasar surveys, rendering the average selection functions of
Fig.~\ref{sdssselfunc} invalid. To illustrate this further, we plot in
Fig.~\ref{sdssselfunc} the average selection function of simulated quasars with
a measurable NUV flux ($m_\mathrm{NUV}<26$ including attenuation by the IGM).
GALEX NUV-detected quasars are unusually blue in $u-g$, and consequently
largely missed by the SDSS color selection criteria.

The inefficiency of SDSS to select high-redshift quasars with blue optical
colors (and likely NUV flux) naturally explains why the \citet{richards06}
selection function substantially overestimates the number counts of NUV-bright
quasars in Fig.~\ref{comparecounts}. Applying instead our color-dependent
selection function lowers the prediction by almost a factor of 2. Unexpectedly,
the predicted number counts are now in excellent agreement with the observed
ones. In total, we predict $\sim 50$ SDSS quasars in the DR7 footprint that can
be detected at S/N$>5$ in the NUV, very close to the actual 52 (58) with
(without) flagging potential cases of source confusion. We predict slightly too
many NUV-bright quasars at $21\la m_\mathrm{NUV}\la 22$, which may be due to the
assumptions regarding the quasar UV spectral energy distribution or the LyC
opacity (the MFP is extrapolated at $z<3.6$, Fig.~\ref{mfpmatch}).

\subsection{The SDSS Lyman limit system bias revisited}

Both, the observed differences in $u-g$ color of color-selected and
radio-selected quasars and the good match of our strongly color-dependent
selection function to observations, point to significant selection effects of
SDSS, either regarding the quasars themselves, or the intergalactic absorption
along their lines of sight. As all relevant spectral parameters of the model
quasars and all IGM absorbers along their sightlines were saved in our MC
simulations, we could explore both possibilities by comparing the statistical
properties of the full MC sample and the subsample fulfilling the SDSS color
selection criteria.

\begin{figure}
\includegraphics[width=\columnwidth]{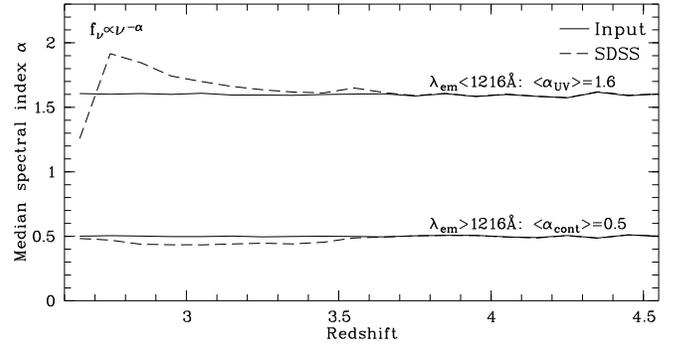}
\caption{\label{sdssseds}Median spectral index $\alpha$ blueward and redward of
\ion{H}{1} Ly$\alpha$ as a function of redshift as adopted in the MC simulations
(solid) and for the subset selected by the SDSS quasar targeting algorithm (dashed).}
\end{figure}

Indeed, we find that the median UV spectral index of SDSS-selected model quasars
is larger at $z_\mathrm{em}<3.5$ (Fig.~\ref{sdssseds}). The Gaussian distribution
of spectral indices is well preserved, but the mean is shifted to higher values,
yielding redder $u-g$ colors. Due to the increasing LyC opacity with redshift
(see below), this bias decreases with increasing redshift. At
$z_\mathrm{em}\simeq 2.7$ there is a sharp break in the UV spectral index
distribution of quasars that would be selected by SDSS. This feature can be
attributed to the inhomogeneities in the SDSS targeting rate in $u-g$ color
space (Fig.~\ref{sdss_u_minus_g}). Blue $u-g$ colors can be due to hard UV
spectral energy distributions, and the different targeting rates may cause
non-trivial changes in the overall appearance of SDSS quasar spectra
(e.g.\ composite spectra) as a function of redshift. The continuum redward of
Ly$\alpha$ is not significantly biased considering our simple model assumptions.
Nevertheless, the slight shift to a harder spectral index at $z_\mathrm{em}<3.5$
might indicate too stringent selection criteria in the other three SDSS colors.
We conclude that SDSS preferentially selects $2.7\la z_\mathrm{em}\la 3.5$
quasars with red spectral energy distributions in the $u$ and $g$ band.

\begin{figure}
\includegraphics[width=\columnwidth]{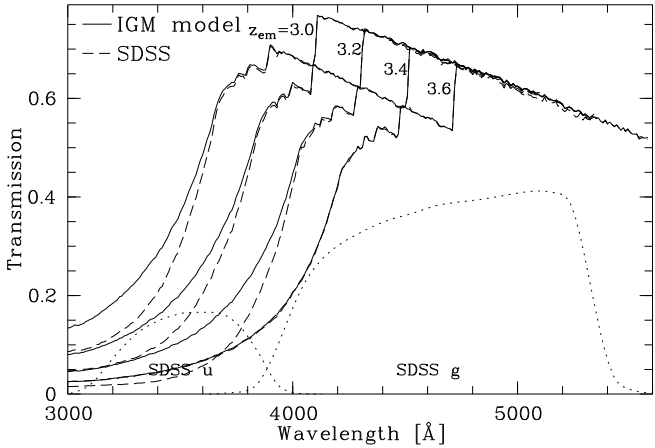}
\caption{\label{sdssigmtrans}Mean IGM Lyman series and continuum transmission
for sources at $z_\mathrm{em}\in\left\lbrace3.0,3.2,3.4,3.6\right\rbrace$ from
our 4000 MC simulations (solid) and for the subset selected by the SDSS quasar
targeting algorithm (dashed). Each sawtooth curve is due to accumulating Lyman
series absorption (first Ly$\alpha$, then Ly$\alpha$+Ly$\beta$, etc.), whereas
the exponential-like roll-off is associated with the LyC opacity. Overplotted
are the SDSS $u$ and $g$ filter curves (dotted).}
\end{figure}

Lyman series and continuum absorption should have an even stronger impact on the
$u-g$ color at these redshifts (Fig.~\ref{examplelos}). Therefore, we computed
the mean IGM Lyman series and continuum transmission at different emission
redshifts, both for the full sample of 4000 MC sightlines and for the subsample
of sightlines towards quasars fulfilling the SDSS color selection criteria in a
$\Delta z_\mathrm{em}=0.02$ window around the emission redshift of interest.
The resulting average 'Lyman valley transmission spectra' \citep[e.g.][]{moller90}
are plotted in Fig.~\ref{sdssigmtrans}. The sample of model spectra is large
enough to clearly show the sawtooth-like features of overlapping Lyman series
absorption. After an initial drop due to beginning series absorption at
$z<z_\mathrm{em}$ the transmission recovers, because high-order high-redshift
absorption overlaps with low-order low-redshift absorption that decreases with
decreasing redshift. Beyond Ly$\epsilon$ there is a quasi-continuous roll-off of
the transmission until LyC absorption sets in. At $z_\mathrm{em}=3.6$ there is
essentially no difference between the average transmission of the full MC sample
and SDSS-selected sightlines (compare the solid and dashed curves). However, at
lower redshifts, the average LyC transmission towards SDSS-selected model
sightlines is much lower than for general sightlines from the MC sample. The on
average stronger LyC absorption corresponds to an on average redder $u-g$ color.
Quasars at these redshifts are still in the vicinity of the stellar locus and
LLSs in their sightlines will significantly redden the $u-g$ color, moving them
away from the stellar locus so that they can be selected by broadband colors.
On the other hand, quasars with little LyC absorption (e.g.\ without LLS) will
have colors similar to main-sequence stars, and are preferentially missed by
broadband color selection. Due to the rarity of LLSs, however, their excess
towards SDSS quasars should not significantly bias the Ly$\alpha$ forest
effective optical depth.

Figure~\ref{sdssigmtrans} presents further evidence that SDSS preferentially
selects sightlines with strong \ion{H}{1} absorbers at $z_\mathrm{em}<3.6$, an
effect that plagued the interpretation of our previous results on the MFP and
the number density of LLSs. In \citet{prochaska09} we found a significant
flattening of the MFP at $z<3.6$ that coincides with an apparent overabundance
of LLSs \citep{prochaska10}. We were puzzled that observed
$z_\mathrm{em}\sim 3.5$ SDSS quasars are significantly redder than their
brethren at $z_\mathrm{em}\sim 3.6$ and suspected that the SDSS color selection
criteria had biased our measurements. The mock quasar spectra processed with
the SDSS color selection routines support our previous claims. The
SDSS-selected model quasars turn redder in $u-g$ towards lower redshifts, as
only these red quasars are outliers from the stellar locus. At
$3\la z_\mathrm{em}\la 3.5$ SDSS is essentially a Lyman break survey, resulting
in an overabundance of LLSs and an underestimated MFP if the analysis is based
solely on quasars from SDSS. Redder UV spectral indices of the quasars can
alleviate this LLS bias somewhat, but not entirely.

\section{Concluding remarks}
\label{conclusions}

We have correlated verified $z_\mathrm{em}>2.7$ quasars to GALEX photometry to
reveal the rare high-redshift quasars whose far-UV fluxes are not extinguished
by intervening Lyman limit systems, with the goal to establish a sample of
UV-bright quasars that likely show intergalactic \ion{He}{2} absorption. We
have used the GALEX UV color $m_\mathrm{FUV}-m_\mathrm{NUV}$ to cull the most
promising targets for follow-up. Red UV colors indicate that the quasar flux is
prematurely truncated redward of the \ion{He}{2} edge, whereas the rare quasars
with blue UV colors and significant FUV flux will likely show flux at
\ion{He}{2} Ly$\alpha$.

We have performed extensive Monte Carlo simulations to estimate the UV color
distribution of UV-bright quasars and their surface density on the sky. We
predict that $\sim 600$ ($\sim 200$) quasars with $z_\mathrm{em}>2.7$ and
$m_{1450}<19$ should be detectable in the NUV (FUV) at $m_\mathrm{UV}<21$
(Fig.~\ref{allskycounts}; without considering sources near the Galactic plane).
The number of UV-bright quasars strongly declines with redshift due to the
declining quasar space density and the increasing \ion{H}{1} Lyman continuum
absorption experienced at the \ion{He}{2} edge (Fig.~\ref{taulycstat}).
Nevertheless, there are enough targets within reach of HST/COS to significantly
constrain \ion{He}{2} reionization by \ion{He}{2} absorption spectra, provided
that the quasars are known and have been imaged with GALEX for efficient pre-selection. 

Most confirmed $z_\mathrm{em}<3.5$ \ion{He}{2} quasars have blue UV colors and
our simulations indicate a $\sim 60$\% \ion{He}{2} detection rate of quasars at
similar UV color (Fig.~\ref{gr4zcolor}), a $\sim$50\% increase over approaches
that do not include color information \citep{syphers09b,syphers09a}. We have
identified 166 additional quasars as prime targets for UV follow-up spectroscopy
with HST/COS to significantly extend the sample of \ion{He}{2} sightlines before
the end of HST's mission (Tables~\ref{candlist1} and \ref{candlist2}).
We have started a survey with HST/COS in Cycle~17 to obtain FUV follow-up spectra
of 8 UV-bright quasars selected from the much smaller GALEX GR3 footprint
(star symbols in Fig.~\ref{gr4zcolor}).

We have reassessed the SDSS color selection efficiency by applying the SDSS
quasar selection criteria to mock photometry of our Monte Carlo spectra. We find
that SDSS preferentially misses UV-bright quasars due to their blue colors that
make them indistinguishable from main-sequence stars (Figs.~\ref{sdssselfunc}
and \ref{sdss_u_minus_g}). The observed $u-g$ colors of color-selected SDSS
quasars are significantly redder than those of radio-selected ones at
$3\la z_\mathrm{em}\la 3.5$, and agree well with our color-dependent SDSS
selection function (Fig.~\ref{sdss_u_minus_g}). These missing quasars lack
strong Lyman continuum absorption due to Lyman limit systems along their lines
of sight that would redden the $u-g$ color (Fig.~\ref{sdssigmtrans}). 

The SDSS color bias has not been well studied previously. Figure~18 of
\citet{bernardi03} reveals that SDSS rarely selected blue quasars at
$3.2<z_\mathrm{em}<3.6$, but the authors did not investigate this further.
\citet{richards06} explored whether primarily radio-selected SDSS quasars have
different color selection efficiencies, but due to low number statistics, they
regarded the differences to be insignificant (their Fig.~10). For the first time
we have been able to demonstrate the full effect and its consequences.

Since the UV-brightest quasars are among the bluest in SDSS $u-g$ at all epochs,
we conclude that SDSS is inefficient in finding further promising targets for
detecting intergalactic helium. Although about two dozen quasars in the SDSS database
already have been confirmed to show \ion{He}{2} \citep{syphers09b,syphers09a},
we predict that the FUV-brightest quasars without strong Lyman continuum absorption
are insensitive to standard color selection techniques.

Due to the restrictive SDSS selection criteria the statistics of high-column
density IGM absorbers measured towards color-selected SDSS quasars will be
biased high \citep{prochaska09,prochaska10}. Our results also indicate that the
incidence of DLAs based on SDSS samples \citep[e.g][]{prochaska05} has been
overestimated. At $3\la z_\mathrm{em}\la 3.5$ the few radio-selected quasars are
probably the only ones within SDSS that are truly unbiased in the statistics of
high-column density absorbers. 

The Lyman limit system bias will also affect the frequency of metal absorption
lines that primarily occur at moderate to high \ion{H}{1} column densities. Because
these absorbers trace the large-scale structure of the IGM, the Lyman limit system
bias might also impact analyses of the clustering properties and the power
spectrum of the Ly$\alpha$ forest. Presumably all $z_\mathrm{em}\ga 3$ quasars
that have been first detected in broadband color surveys are affected by such a
bias to some sort, depending on the exact color selection criteria and the number
and effective wavelengths of the employed filters. Because the abundance of
optically thick absorbers is far less constrained in reality than in
'completeness' simulations like ours, determinations of the optical quasar
luminosity function at $3<z_\mathrm{em}<3.5$ should invoke a variety of IGM
models which will result in large systematic uncertainties in the luminosity
function. Our study indicates that results based on color-selected high-redshift
quasar samples are not as easy to interpret as previously thought due to the
intertwined demographics of strong IGM absorbers and their background candles.

\acknowledgments
We thank Gordon Richards for running our mock photometry through the SDSS quasar
target selection routine. Robert da~Silva kindly took the spectrum of
J1943$-$1502 presented in Fig.~\ref{sypherscand}. We acknowledge support by an
NSF CAREER grant (AST-0548180) and by NSF grant AST-0908910. Further support
for this work was provided by NASA through grant number GO-11742 from the
Space Telescope Science Institute, which is operated by the Association of
Universities for Research in Astronomy, Inc., under NASA contract NAS 5-26555.

GALEX (Galaxy Evolution Explorer) is a NASA Small Explorer. We acknowledge NASA's
support for construction, operation, and science analysis for the GALEX mission,
developed in corporation with the Centre National d?Etudes Spatiales of France
and the Korean Ministry of Science and Technology. GALEX is operated for NASA by
the California Institute of Technology under NASA contract NAS5-98034. 
Based on observations made with the NASA/ESA Hubble Space Telescope, obtained
from the data archive at STScI.

This research has made use of the NASA/IPAC Extragalactic Database (NED) which
is operated by the Jet Propulsion Laboratory, California Institute of Technology,
under contract with the National Aeronautics and Space Administration.

Funding for the SDSS and SDSS-II has been provided by the Alfred P. Sloan Foundation,
the Participating Institutions, the National Science Foundation, the U.S. Department
of Energy, the National Aeronautics and Space Administration, the Japanese
Monbukagakusho, the Max Planck Society, and the Higher Education Funding Council
for England. The SDSS Web Site is http://www.sdss.org/.
The SDSS is managed by the Astrophysical Research Consortium for the Participating
Institutions. The Participating Institutions are the American Museum of Natural
History, Astrophysical Institute Potsdam, University of Basel, University of
Cambridge, Case Western Reserve University, University of Chicago, Drexel University,
Fermilab, the Institute for Advanced Study, the Japan Participation Group,
Johns Hopkins University, the Joint Institute for Nuclear Astrophysics, the Kavli
Institute for Particle Astrophysics and Cosmology, the Korean Scientist Group,
the Chinese Academy of Sciences (LAMOST), Los Alamos National Laboratory,
the Max-Planck-Institute for Astronomy (MPIA), the Max-Planck-Institute for
Astrophysics (MPA), New Mexico State University, Ohio State University,
University of Pittsburgh, University of Portsmouth, Princeton University,
the United States Naval Observatory, and the University of Washington.

{{\it Facilities:} \facility{GALEX}, \facility{HST}}.

\bibliography{he2galex}

\clearpage
\tabletypesize{\footnotesize}
\begin{deluxetable*}{lccccccccrrc}
\tablewidth{0pt}
\tablecaption{\label{candlist1}Promising targets to search for \ion{He}{2} (GALEX S/N$>3$ and $m_\mathrm{FUV}-m_\mathrm{NUV}<1$).}
\tablehead{
\colhead{Object}&\colhead{$\alpha$ (J2000)}&\colhead{$\delta$ (J2000)}&\colhead{$z_\mathrm{em}$}&\colhead{$m_\mathrm{opt}$\tablenotemark{a}}&\colhead{filter}	&\colhead{$m_\mathrm{FUV}$\,[AB]}&\colhead{$m_\mathrm{NUV}$\,[AB]}&\colhead{limit\tablenotemark{b}}&\colhead{S/N$_\mathrm{FUV}$}&\colhead{S/N$_\mathrm{NUV}$}&\colhead{neighbors\tablenotemark{c}}
}
\startdata
2QZ~J1411$-$0229	&$14^\mathrm{h}11^\mathrm{m}24\fs63$	&$-02\degr29\arcmin42\farcs6$	&$2.702$	&$19.42$	&$r$	&$22.52$	&$21.76$	&0	&5.3	&9.1	&0\\
FIRST~J1007$+$4003	&$10^\mathrm{h}07^\mathrm{m}16\fs85$	&$+40\degr03\arcmin56\farcs3$	&$2.728$	&$19.11$	&$r$	&$20.33$	&$20.65$	&0	&4.6	&5.3	&0\\
SDSS~J2330$+$0001	&$23^\mathrm{h}30^\mathrm{m}26\fs26$	&$+00\degr01\arcmin23\farcs9$	&$2.733$	&$20.04$	&$r$	&$23.79$	&$24.79$	&0	&12.3	&5.5	&3\\
CTS~0216		&$02^\mathrm{h}16^\mathrm{m}23\fs10$	&$-39\degr07\arcmin55\farcs0$	&$2.735$	&$17.90$	&$B$	&$19.77$	&$19.37$	&0	&5.2	&11.9	&2\\
SDSS~J0029$+$0019	&$00^\mathrm{h}29^\mathrm{m}12\fs91$	&$+00\degr19\arcmin46\farcs6$	&$2.736$	&$18.66$	&$r$	&$20.68$	&$20.70$	&0	&25.6	&28.7	&1\\
SDSS~J0142$-$0027	&$01^\mathrm{h}42^\mathrm{m}43\fs54$	&$-00\degr27\arcmin54\farcs0$	&$2.737$	&$19.93$	&$r$	&$22.85$	&$22.58$	&0	&4.1	&5.2	&0\\
SDSS~J2358$-$0032	&$23^\mathrm{h}58^\mathrm{m}07\fs79$	&$-00\degr32\arcmin24\farcs5$	&$2.753$	&$19.14$	&$r$	&$21.67$	&$21.31$	&0	&9.9	&16.7	&0\\
UM~682			&$03^\mathrm{h}10^\mathrm{m}28\fs10$	&$-19\degr09\arcmin43\farcs7$	&$2.756$	&$17.90$	&$V$	&$19.68$	&$19.65$	&0	&7.2	&11.6	&2\\
PC~2204$+$0127		&$22^\mathrm{h}06^\mathrm{m}46\fs19$	&$+01\degr41\arcmin45\farcs7$	&$2.757$	&$19.07$	&$R$	&$22.33$	&$21.43$	&0	&5.0	&12.0	&2\\
PC~1640$+$4711		&$16^\mathrm{h}41^\mathrm{m}25\fs86$	&$+47\degr05\arcmin45\farcs8$	&$2.770$	&$19.51$	&$r$	&$23.20$	&$23.06$	&0	&4.0	&4.8	&0\\
SDSS~J2324$-$0005	&$23^\mathrm{h}24^\mathrm{m}52\fs55$	&$-00\degr05\arcmin15\farcs3$	&$2.779$	&$19.43$	&$r$	&$22.31$	&$22.16$	&0	&31.1	&32.9	&3\\
SDSS~J1309$-$0333	&$13^\mathrm{h}09^\mathrm{m}34\fs18$	&$-03\degr33\arcmin18\farcs4$	&$2.781$	&$19.19$	&$r$	&$22.55$	&$22.24$	&0	&5.6	&6.8	&0\\
SDSS~J0809$+$3116	&$08^\mathrm{h}09^\mathrm{m}12\fs68$	&$+31\degr16\arcmin02\farcs1$	&$2.796$	&$18.86$	&$r$	&$21.27$	&$20.97$	&0	&3.1	&6.7	&0\\
Q~0207$-$398		&$02^\mathrm{h}09^\mathrm{m}28\fs59$	&$-39\degr39\arcmin39\farcs5$	&$2.805$	&$17.15$	&$V$	&$20.29$	&$21.07$	&0	&5.2	&4.2	&2\\
LBQS~1216$+$1656	&$12^\mathrm{h}19^\mathrm{m}20\fs40$	&$+16\degr39\arcmin29\farcs5$	&$2.818$	&$18.17$	&$r$	&$20.80$	&$20.04$	&0	&3.5	&6.9	&0\\
SDSS~J1519$+$3609	&$15^\mathrm{h}19^\mathrm{m}10\fs37$	&$+36\degr09\arcmin40\farcs5$	&$2.819$	&$18.70$	&$r$	&$20.87$	&$20.59$	&0	&5.2	&9.5	&0\\
Q~2315$-$4230		&$23^\mathrm{h}18^\mathrm{m}15\fs10$	&$-42\degr13\arcmin48\farcs0$	&$2.830$	&$20.00$	&$V$	&$21.94$	&$21.37$	&0	&11.6	&21.0	&2\\
SDSS~J1230$-$0253	&$12^\mathrm{h}30^\mathrm{m}53\fs16$	&$-02\degr53\arcmin52\farcs0$	&$2.837$	&$18.92$	&$r$	&$21.88$	&$23.30$	&0	&7.0	&3.3	&0\\
2QZ~J2158$-$3037	&$21^\mathrm{h}58^\mathrm{m}29\fs66$	&$-30\degr37\arcmin21\farcs6$	&$2.838$	&$20.35$	&$b_J$	&$24.57$	&$24.61$	&1	&3.9	&3.3	&3\\
HS~1024$+$1849		&$10^\mathrm{h}27^\mathrm{m}34\fs13$	&$+18\degr34\arcmin27\farcs5$	&$2.840$	&$17.83$	&$r$	&$19.97$	&$19.82$	&0	&5.3	&16.9	&0\\
SDSS~J0141$+$1341	&$01^\mathrm{h}41^\mathrm{m}34\fs01$	&$+13\degr41\arcmin58\farcs9$	&$2.843$	&$19.52$	&$r$	&$22.95$	&$22.76$	&0	&3.5	&5.8	&0\\
SDSS~J2156$+$0037	&$21^\mathrm{h}56^\mathrm{m}04\fs18$	&$+00\degr37\arcmin42\farcs3$	&$2.844$	&$19.01$	&$r$	&$21.90$	&$21.39$	&0	&9.2	&13.3	&1\\
CSO~0806		&$13^\mathrm{h}04^\mathrm{m}11\fs99$	&$+29\degr53\arcmin48\farcs8$	&$2.850$	&$17.65$	&$r$	&$20.46$	&$20.47$	&0	&15.9	&26.1	&0\\
SDSS~J2331$+$0036	&$23^\mathrm{h}31^\mathrm{m}31\fs48$	&$+00\degr36\arcmin44\farcs4$	&$2.852$	&$19.53$	&$r$	&$22.55$	&$21.96$	&0	&4.1	&5.3	&0\\
SBS~1602$+$576		&$16^\mathrm{h}03^\mathrm{m}55\fs92$	&$+57\degr30\arcmin54\farcs4$	&$2.858$	&$17.33$	&$r$	&$19.76$	&$19.08$	&0	&8.2	&16.1	&0\\
PMN~J1404$+$0728	&$14^\mathrm{h}04^\mathrm{m}32\fs99$	&$+07\degr28\arcmin46\farcs9$	&$2.866$	&$18.87$	&$r$	&$20.78$	&$21.29$	&0	&4.0	&3.7	&0\\
PC~0058$+$0215		&$01^\mathrm{h}00^\mathrm{m}58\fs40$	&$+02\degr31\arcmin32\farcs0$	&$2.868$	&$18.91$	&$R$	&$21.32$	&$21.22$	&0	&6.4	&7.4	&2\\
CTS~0347		&$22^\mathrm{h}05^\mathrm{m}36\fs26$	&$-34\degr26\arcmin03\farcs9$	&$2.870$	&$18.70$	&$R$	&$20.89$	&$20.48$	&0	&5.4	&7.9	&2\\
2QZ~J0126$-$3124	&$01^\mathrm{h}26^\mathrm{m}00\fs17$	&$-31\degr24\arcmin21\farcs5$	&$2.881$	&$20.43$	&$b_J$	&$21.48$	&$22.84$	&0	&4.6	&3.0	&2\\
SDSS~J1626$+$3856	&$16^\mathrm{h}26^\mathrm{m}12\fs99$	&$+38\degr56\arcmin27\farcs2$	&$2.882$	&$18.63$	&$r$	&$21.23$	&$21.82$	&0	&3.2	&3.1	&0\\
SDSS~J2342$-$0042	&$23^\mathrm{h}42^\mathrm{m}36\fs90$	&$-00\degr42\arcmin32\farcs8$	&$2.885$	&$20.47$	&$r$	&$24.06$	&$23.21$	&0	&4.6	&4.2	&0\\
SDSS~J1410$+$4727	&$14^\mathrm{h}10^\mathrm{m}59\fs61$	&$+47\degr27\arcmin33\farcs3$	&$2.901$	&$19.38$	&$r$	&$21.74$	&$21.31$	&0	&3.1	&5.2	&0\\
SDSS~J1443$+$3546	&$14^\mathrm{h}43^\mathrm{m}11\fs58$	&$+35\degr46\arcmin46\farcs3$	&$2.941$	&$18.79$	&$r$	&$20.96$	&$21.19$	&0	&4.3	&4.6	&0\\
RDS~477A		&$10^\mathrm{h}53^\mathrm{m}06\fs04$	&$+57\degr34\arcmin24\farcs6$	&$2.949$	&$20.47$	&$r$	&$24.55$	&$24.51$	&0	&3.1	&3.6	&3\\
SDSS~J0818$+$4908	&$08^\mathrm{h}18^\mathrm{m}50\fs01$	&$+49\degr08\arcmin17\farcs0$	&$2.954$	&$18.52$	&$r$	&$21.49$	&$21.46$	&0	&11.6	&22.6	&0\\
SDSS~J1033$+$5406	&$10^\mathrm{h}33^\mathrm{m}10\fs71$	&$+54\degr06\arcmin46\farcs8$	&$2.959$	&$19.27$	&$r$	&$22.55$	&$23.13$	&0	&4.9	&4.4	&0\\
FIRST~J1456$-$0218	&$14^\mathrm{h}56^\mathrm{m}40\fs98$	&$-02\degr18\arcmin19\farcs4$	&$2.963$	&$19.53$	&$r$	&$22.80$	&$22.07$	&0	&4.5	&7.4	&0\\
SDSS~J0922$+$5321	&$09^\mathrm{h}22^\mathrm{m}47\fs83$	&$+53\degr21\arcmin46\farcs6$	&$3.000$	&$19.75$	&$r$	&$22.76$	&$23.32$	&0	&4.1	&3.4	&0\\
SDSS~J1657$+$3553	&$16^\mathrm{h}57^\mathrm{m}51\fs68$	&$+35\degr53\arcmin18\farcs0$	&$3.005$	&$19.23$	&$r$	&$23.35$	&$22.81$	&0	&5.2	&7.4	&0\\
SDSS~J0905$+$3057	&$09^\mathrm{h}05^\mathrm{m}08\fs88$	&$+30\degr57\arcmin57\farcs3$	&$3.027$	&$17.37$	&$r$	&$20.89$	&$21.62$	&0	&6.5	&5.5	&0\\
SDSS~J1101$+$1053	&$11^\mathrm{h}01^\mathrm{m}55\fs73$	&$+10\degr53\arcmin02\farcs3$	&$3.031$	&$18.97$	&$r$	&$21.59$	&$21.94$	&0	&4.0	&4.0	&0\\
SDSS~J1244$+$6201	&$12^\mathrm{h}44^\mathrm{m}56\fs98$	&$+62\degr01\arcmin43\farcs0$	&$3.057$	&$18.63$	&$r$	&$21.08$	&$21.33$	&0	&4.7	&7.8	&0\\
SDSS~J1052$+$2543	&$10^\mathrm{h}52^\mathrm{m}54\fs49$	&$+25\degr43\arcmin03\farcs9$	&$3.062$	&$18.52$	&$r$	&$21.43$	&$20.96$	&0	&3.1	&3.9	&0\\
PC~2211$+$0119		&$22^\mathrm{h}14^\mathrm{m}27\fs81$	&$+01\degr34\arcmin57\farcs3$	&$3.100$	&$19.10$	&$R$	&$22.26$	&$22.35$	&0	&5.4	&6.7	&2\\
SDSS~J1025$+$0452	&$10^\mathrm{h}25^\mathrm{m}09\fs63$	&$+04\degr52\arcmin46\farcs7$	&$3.244$	&$18.02$	&$r$	&$21.37$	&$21.72$	&0	&3.8	&3.1	&0\\
SDSS~J0955$+$6842	&$09^\mathrm{h}55^\mathrm{m}54\fs30$	&$+68\degr42\arcmin01\farcs2$	&$3.269$	&$19.26$	&$r$	&$24.05$	&$24.06$	&0	&5.4	&6.1	&3\\
SDSS~J1220$+$4549	&$12^\mathrm{h}20^\mathrm{m}17\fs06$	&$+45\degr49\arcmin41\farcs1$	&$3.293$	&$18.20$	&$r$	&$22.78$	&$22.83$	&0	&3.8	&5.7	&0\\
HS~0911$+$4809		&$09^\mathrm{h}15^\mathrm{m}10\fs01$	&$+47\degr56\arcmin58\farcs7$	&$3.337$	&$17.84$	&$r$	&$20.53$	&$20.25$	&0	&5.3	&9.4	&0\\
SDSS~J0054$+$0028	&$00^\mathrm{h}54^\mathrm{m}01\fs48$	&$+00\degr28\arcmin47\farcs7$	&$3.413$	&$19.93$	&$r$	&$22.06$	&$21.79$	&0	&7.6	&8.6	&1\\
CLASXS~449		&$10^\mathrm{h}34^\mathrm{m}58\fs01$	&$+57\degr50\arcmin46\farcs5$	&$3.430$	&$23.80$	&$R$	&$24.19$	&$23.63$	&0	&4.5	&6.1	&0\\
SDSS~J1233$+$0941	&$12^\mathrm{h}33^\mathrm{m}02\fs74$	&$+09\degr41\arcmin44\farcs2$	&$3.816$	&$20.36$	&$r$	&$23.66$	&$23.03$	&0	&3.7	&3.4	&0\\
CDFN~097		&$12^\mathrm{h}36^\mathrm{m}12\fs93$	&$+62\degr19\arcmin29\farcs8$	&$3.938$	&$22.80$	&$R$	&$25.78$	&$24.97$	&0	&4.5	&5.3	&0
\enddata
\tablecomments{39 of the 41 $z_\mathrm{em}>2.78$ quasars were previously suggested as candidate \ion{He}{2} quasars by \citet{syphers09b}.}
\tablenotetext{a}{SDSS $r$ AB magnitude if filter is $r$, otherwise Vega magnitude in given filter.}
\tablenotetext{b}{GALEX limit flag. 0: formal two-band detection, 1: $1\sigma$ lower limit in $m_\mathrm{FUV}$, 2: $1\sigma$ lower limit in $m_\mathrm{NUV}$}
\tablenotetext{c}{Neighbor flag. 0: no SDSS source within $r<5\arcsec$ of the quasar, 1: sufficiently red SDSS source within $r<5\arcsec$ of the quasar, 2: quasar not imaged in SDSS DR7, 3: potential source confusion (DIS detection)}
\end{deluxetable*}

\clearpage
\LongTables
\begin{deluxetable*}{lccccccccrrc}
\tablecolumns{12}
\tablecaption{\label{candlist2}Further quasars with potential FUV flux (GALEX S/N$_\mathrm{FUV}>2$).\tablenotemark{a}}
\tablewidth{0pt}
\tablehead{
\colhead{Object}&\colhead{$\alpha$ (J2000)}&\colhead{$\delta$ (J2000)}&\colhead{$z_\mathrm{em}$}&\colhead{$m_\mathrm{opt}$\tablenotemark{b}}&\colhead{filter}	&\colhead{$m_\mathrm{FUV}$\,[AB]}&\colhead{$m_\mathrm{NUV}$\,[AB]}&\colhead{limit\tablenotemark{c}}&\colhead{S/N$_\mathrm{FUV}$}&\colhead{S/N$_\mathrm{NUV}$}&\colhead{neighbors\tablenotemark{d}}
}
\startdata
2QZ~J0035$-$2837	&$00^\mathrm{h}35^\mathrm{m}24\fs23$	&$-28\degr37\arcmin14\farcs7$	&$2.702$	&$20.73$	&$b_J$	&$22.70$	&$22.27$	&0	&2.2	&2.6	&2\\
2QZ~J0258$-$2941	&$02^\mathrm{h}58^\mathrm{m}09\fs15$	&$-29\degr41\arcmin08\farcs7$	&$2.702$	&$20.12$	&$b_J$	&$21.78$	&$22.43$	&0	&3.5	&2.8	&2\\
SDSS~J1039$+$3040	&$10^\mathrm{h}39^\mathrm{m}24\fs05$	&$+30\degr40\arcmin59\farcs5$	&$2.705$	&$19.99$	&$r$	&$21.96$	&$23.78$	&2	&2.9	&0.7	&0\\
SDSS~J1301$-$0038	&$13^\mathrm{h}01^\mathrm{m}47\fs88$	&$-00\degr38\arcmin17\farcs3$	&$2.705$	&$19.40$	&$r$	&$22.99$	&$24.47$	&2	&2.4	&0	&0\\
2QZ~J2153$-$2719	&$21^\mathrm{h}53^\mathrm{m}16\fs08$	&$-27\degr19\arcmin38\farcs6$	&$2.706$	&$20.04$	&$b_J$	&$22.14$	&$21.59$	&0	&2.5	&3.8	&2\\
2QZ~J0203$-$3153	&$02^\mathrm{h}03^\mathrm{m}15\fs58$	&$-31\degr53\arcmin54\farcs9$	&$2.710$	&$20.65$	&$b_J$	&$22.21$	&$20.95$	&2	&2.5	&2.0	&2\\
CTS~0538		&$14^\mathrm{h}21^\mathrm{m}01\fs60$	&$-23\degr07\arcmin32\farcs0$	&$2.710$	&$18.50$	&$R$	&$21.31$	&$21.35$	&0	&2.7	&3.0	&2\\
SDSS~J1407$+$2127	&$14^\mathrm{h}07^\mathrm{m}01\fs12$	&$+21\degr27\arcmin15\farcs9$	&$2.711$	&$18.35$	&$r$	&$21.86$	&$20.88$	&0	&2.2	&6.8	&0\\
SDSS~J1325$+$0814	&$13^\mathrm{h}25^\mathrm{m}17\fs85$	&$+08\degr14\arcmin08\farcs4$	&$2.715$	&$18.70$	&$r$	&$22.31$	&$22.69$	&0	&2.8	&2.5	&0\\
SDSS~J0014$-$0112	&$00^\mathrm{h}14^\mathrm{m}43\fs69$	&$-01\degr12\arcmin06\farcs4$	&$2.717$	&$18.84$	&$r$	&$23.99$	&$21.76$	&0	&3.7	&12.3	&0\\
Q~0040$-$370		&$00^\mathrm{h}42^\mathrm{m}43\fs93$	&$-36\degr47\arcmin41\farcs5$	&$2.723$	&$17.85$	&$V$	&$21.30$	&$21.33$	&0	&2.8	&3.8	&2\\
2QZ~J0141$-$3209	&$01^\mathrm{h}41^\mathrm{m}54\fs69$	&$-32\degr09\arcmin11\farcs6$	&$2.724$	&$20.03$	&$b_J$	&$22.77$	&$22.20$	&2	&2.3	&4.3	&2\\
SDSS~J1159$+$0222	&$11^\mathrm{h}59^\mathrm{m}04\fs30$	&$+02\degr22\arcmin14\farcs1$	&$2.725$	&$19.11$	&$r$	&$22.40$	&$21.10$	&1	&3.5	&10.7	&0\\
Q~1613$+$172		&$16^\mathrm{h}15^\mathrm{m}56\fs87$	&$+17\degr07\arcmin51\farcs4$	&$2.729$	&$18.24$	&$r$	&$22.00$	&$22.78$	&0	&2.8	&2.5	&0\\
QSO~J0059$-$3541	&$00^\mathrm{h}59^\mathrm{m}14\fs21$	&$-35\degr41\arcmin42\farcs1$	&$2.730$	&$18.04$	&$V$	&$22.24$	&$20.97$	&0	&6.3	&16.4	&2\\
SDSS~J1026$+$2842	&$10^\mathrm{h}26^\mathrm{m}54\fs39$	&$+28\degr42\arcmin54\farcs5$	&$2.739$	&$19.68$	&$r$	&$22.62$	&$22.09$	&0	&2.2	&3.7	&0\\
HE~0151$-$4326		&$01^\mathrm{h}53^\mathrm{m}27\fs20$	&$-43\degr11\arcmin38\farcs0$	&$2.740$	&$17.19$	&$b_J$	&$20.63$	&$19.30$	&0	&5.8	&15.5	&2\\
2QZ~J1129$+$0134	&$11^\mathrm{h}29^\mathrm{m}57\fs65$	&$+01\degr34\arcmin16\farcs0$	&$2.743$	&$19.74$	&$r$	&$22.35$	&$22.49$	&1	&2.2	&2.7	&1\\
2QZ~J1326$+$0042	&$13^\mathrm{h}26^\mathrm{m}22\fs41$	&$+00\degr42\arcmin37\farcs3$	&$2.743$	&$18.72$	&$r$	&$21.66$	&$21.65$	&0	&2.7	&3.4	&0\\
2QZ~J0053$-$3140	&$00^\mathrm{h}53^\mathrm{m}30\fs68$	&$-31\degr40\arcmin18\farcs8$	&$2.751$	&$19.99$	&$b_J$	&$23.67$	&$21.87$	&0	&2.1	&7.1	&2\\
2QZ~J0012$-$3131	&$00^\mathrm{h}12^\mathrm{m}43\fs11$	&$-31\degr31\arcmin13\farcs6$	&$2.755$	&$19.84$	&$b_J$	&$22.64$	&$22.35$	&1	&2.2	&3.2	&2\\
HELLAS~149		&$20^\mathrm{h}44^\mathrm{m}34\fs80$	&$-10\degr28\arcmin08\farcs0$	&$2.755$	&$17.79$	&$V$	&$21.34$	&$20.15$	&0	&3.7	&8.0	&2\\
QSO~J0056$-$4013	&$00^\mathrm{h}56^\mathrm{m}11\fs76$	&$-40\degr13\arcmin16\farcs2$	&$2.758$	&$18.10$	&$R$	&$21.58$	&$23.15$	&0	&2.7	&2.2	&2\\
SDSS~J1600$+$4033	&$16^\mathrm{h}00^\mathrm{m}33\fs09$	&$+40\degr33\arcmin43\farcs9$	&$2.761$	&$19.20$	&$r$	&$22.28$	&$21.58$	&1	&2.4	&6.1	&0\\
SDSS~J0150$-$0825	&$01^\mathrm{h}50^\mathrm{m}09\fs46$	&$-08\degr25\arcmin10\farcs8$	&$2.763$	&$18.96$	&$r$	&$23.64$	&$23.29$	&0	&2.7	&3.4	&0\\
SDSS~J0809$+$0658	&$08^\mathrm{h}09^\mathrm{m}46\fs14$	&$+06\degr58\arcmin07\farcs9$	&$2.763$	&$20.04$	&$r$	&$24.17$	&$24.01$	&1	&2.1	&2.4	&0\\
SDSS~J1546$+$2315	&$15^\mathrm{h}46^\mathrm{m}59\fs33$	&$+23\degr15\arcmin47\farcs3$	&$2.777$	&$17.80$	&$r$	&$22.22$	&$22.65$	&1	&2.3	&2.9	&0\\
2QZ~J0034$-$3048	&$00^\mathrm{h}34^\mathrm{m}47\fs21$	&$-30\degr48\arcmin13\farcs5$	&$2.785$	&$19.97$	&$b_J$	&$22.59$	&$21.46$	&0	&2.7	&6.4	&2\\
SDSS~J1418$+$5858	&$14^\mathrm{h}18^\mathrm{m}22\fs89$	&$+58\degr58\arcmin06\farcs4$	&$2.785$	&$17.78$	&$r$	&$21.74$	&$19.94$	&0	&2.3	&11.7	&0\\
LBQS~0041$-$2707	&$00^\mathrm{h}43^\mathrm{m}51\fs83$	&$-26\degr51\arcmin27\farcs5$	&$2.786$	&$17.83$	&$V$	&$21.79$	&$22.01$	&0	&3.3	&3.0	&2\\
2QZ~J0044$-$3147	&$00^\mathrm{h}44^\mathrm{m}05\fs04$	&$-31\degr47\arcmin04\farcs5$	&$2.789$	&$19.80$	&$b_J$	&$22.66$	&$22.27$	&0	&2.7	&5.5	&2\\
2QZ~J2223$-$3131	&$22^\mathrm{h}23^\mathrm{m}12\fs45$	&$-31\degr31\arcmin29\farcs4$	&$2.792$	&$19.44$	&$b_J$	&$22.07$	&$22.59$	&0	&2.8	&2.8	&2\\
SDSS~J0103$+$0026	&$01^\mathrm{h}03^\mathrm{m}37\fs46$	&$+00\degr26\arcmin08\farcs2$	&$2.795$	&$20.35$	&$r$	&$23.54$	&$25.65$	&2	&4.4	&0	&0\\
2QZ~J1428$+$0010	&$14^\mathrm{h}28^\mathrm{m}49\fs85$	&$+00\degr10\arcmin40\farcs7$	&$2.807$	&$19.79$	&$r$	&$21.65$	&$23.60$	&2	&2.2	&0	&0\\
H~0853$+$1953		&$08^\mathrm{h}56^\mathrm{m}26\fs47$	&$+19\degr41\arcmin37\farcs7$	&$2.818$	&$18.74$	&$r$	&$23.31$	&$22.17$	&0	&3.8	&6.5	&0\\
SDSS~J0225$+$0048	&$02^\mathrm{h}25^\mathrm{m}19\fs50$	&$+00\degr48\arcmin23\farcs6$	&$2.820$	&$20.54$	&$r$	&$24.82$	&$23.01$	&0	&2.4	&5.7	&0\\
SDSS~J0030$+$0053	&$00^\mathrm{h}30^\mathrm{m}17\fs11$	&$+00\degr53\arcmin58\farcs8$	&$2.831$	&$19.92$	&$r$	&$23.78$	&$24.34$	&0	&3.3	&2.4	&0\\
FIRST~J0905$+$3555	&$09^\mathrm{h}05^\mathrm{m}36\fs07$	&$+35\degr55\arcmin51\farcs6$	&$2.839$	&$18.39$	&$r$	&$21.88$	&$21.60$	&0	&2.3	&3.2	&0\\
SDSS~J1504$-$0008	&$15^\mathrm{h}04^\mathrm{m}25\fs53$	&$-00\degr08\arcmin03\farcs2$	&$2.840$	&$18.92$	&$r$	&$22.44$	&$23.94$	&2	&3.9	&0.9	&0\\
2QZ~J0024$-$3149	&$00^\mathrm{h}24^\mathrm{m}16\fs22$	&$-31\degr49\arcmin42\farcs9$	&$2.846$	&$20.24$	&$b_J$	&$22.56$	&$22.36$	&0	&2.6	&3.0	&2\\
UM~658			&$22^\mathrm{h}46^\mathrm{m}52\fs66$	&$-22\degr03\arcmin09\farcs2$	&$2.852$	&$17.80$	&$V$	&$22.40$	&$21.80$	&0	&2.3	&2.5	&2\\
SDSS~J0034$-$0109	&$00^\mathrm{h}34^\mathrm{m}20\fs62$	&$-01\degr09\arcmin17\farcs3$	&$2.854$	&$20.24$	&$r$	&$23.73$	&$22.55$	&0	&3.6	&6.5	&0\\
SDSS~J1309$+$2815	&$13^\mathrm{h}09^\mathrm{m}39\fs49$	&$+28\degr15\arcmin08\farcs0$	&$2.854$	&$18.99$	&$r$	&$21.71$	&$23.17$	&2	&2.7	&0.1	&0\\
SDSS~J1439$+$0421	&$14^\mathrm{h}39^\mathrm{m}48\fs06$	&$+04\degr21\arcmin12\farcs8$	&$2.857$	&$19.00$	&$r$	&$23.99$	&$24.16$	&0	&2.6	&2.2	&0\\
SDSS~J1241$+$2719	&$12^\mathrm{h}41^\mathrm{m}40\fs98$	&$-27\degr19\arcmin27\farcs5$	&$2.862$	&$19.21$	&$r$	&$22.54$	&$23.86$	&2	&2.0	&1.1	&0\\
SDSS~J0039$+$1527	&$00^\mathrm{h}39^\mathrm{m}39\fs96$	&$+15\degr27\arcmin20\farcs3$	&$2.867$	&$19.14$	&$r$	&$23.04$	&$23.99$	&2	&4.0	&1.9	&0\\
FIRST~J1231$+$0102	&$12^\mathrm{h}31^\mathrm{m}39\fs12$	&$+01\degr02\arcmin29\farcs3$	&$2.883$	&$18.33$	&$r$	&$22.67$	&$20.57$	&0	&3.9	&26.2	&0\\
SDSS~J1154$+$4030	&$11^\mathrm{h}54^\mathrm{m}13\fs87$	&$+40\degr30\arcmin00\farcs1$	&$2.893$	&$20.36$	&$r$	&$21.64$	&$23.27$	&2	&2.4	&1.0	&0\\
SDSS~J0130$-$0007	&$01^\mathrm{h}30^\mathrm{m}43\fs41$	&$-00\degr07\arcmin35\farcs3$	&$2.894$	&$19.95$	&$r$	&$23.76$	&$23.57$	&0	&2.8	&2.2	&0\\
2QZ~J0114$-$2719	&$01^\mathrm{h}14^\mathrm{m}19\fs16$	&$-27\degr19\arcmin12\farcs4$	&$2.896$	&$20.55$	&$b_J$	&$22.64$	&$23.30$	&2	&2.6	&2.1	&2\\
SDSS~J1322$+$3955	&$13^\mathrm{h}22^\mathrm{m}59\fs97$	&$+39\degr55\arcmin29\farcs9$	&$2.898$	&$18.35$	&$r$	&$22.19$	&$23.10$	&0	&3.4	&2.7	&0\\
SDSS~J1427$+$0014	&$14^\mathrm{h}27^\mathrm{m}09\fs81$	&$+00\degr14\arcmin50\farcs2$	&$2.908$	&$18.54$	&$r$	&$23.56$	&$23.32$	&0	&2.6	&2.8	&0\\
PKS~0246$-$231		&$02^\mathrm{h}48^\mathrm{m}22\fs74$	&$-22\degr57\arcmin58\farcs2$	&$2.914$	&$20.00$	&$R$	&$22.15$	&$21.42$	&0	&2.3	&3.1	&2\\
SDSS~J1525$+$2207	&$15^\mathrm{h}25^\mathrm{m}34\fs50$	&$+22\degr07\arcmin00\farcs7$	&$2.914$	&$19.12$	&$r$	&$21.92$	&$21.87$	&0	&2.6	&2.8	&1\\
SDSS~J1210$+$3509	&$12^\mathrm{h}10^\mathrm{m}40\fs36$	&$+35\degr09\arcmin11\farcs3$	&$2.919$	&$19.87$	&$r$	&$22.59$	&$21.72$	&1	&2.1	&5.1	&0\\
FIRST~J0936$+$2927	&$09^\mathrm{h}36^\mathrm{m}43\fs51$	&$+29\degr27\arcmin13\farcs6$	&$2.926$	&$18.11$	&$r$	&$20.80$	&$20.59$	&0	&3.0	&4.7	&0\\
SDSS~J0300$-$0749	&$03^\mathrm{h}00^\mathrm{m}47\fs62$	&$-07\degr49\arcmin02\farcs8$	&$2.939$	&$20.02$	&$r$	&$22.84$	&$21.73$	&0	&3.7	&4.3	&0\\
FIRST~J1604$+$1645	&$16^\mathrm{h}04^\mathrm{m}41\fs47$	&$+16\degr45\arcmin38\farcs3$	&$2.939$	&$16.68$	&$r$	&$21.01$	&$19.56$	&0	&4.7	&16.2	&0\\
FIRST~J1159$+$4136	&$11^\mathrm{h}59^\mathrm{m}47\fs10$	&$+41\degr36\arcmin59\farcs1$	&$2.944$	&$18.71$	&$r$	&$22.12$	&$21.93$	&0	&2.6	&4.2	&0\\
FIRST~J1332$+$0805	&$13^\mathrm{h}32^\mathrm{m}18\fs55$	&$+08\degr05\arcmin48\farcs3$	&$2.947$	&$18.86$	&$r$	&$21.90$	&$23.73$	&2	&2.6	&0	&1\\
SDSS~J0905$+$4107	&$09^\mathrm{h}05^\mathrm{m}18\fs02$	&$+41\degr07\arcmin57\farcs6$	&$2.954$	&$19.70$	&$r$	&$22.56$	&$23.01$	&1	&2.0	&1.8	&1\\
QSO~J1334$+$2801	&$13^\mathrm{h}34^\mathrm{m}36\fs63$	&$+28\degr01\arcmin41\farcs5$	&$2.958$	&$19.17$	&$r$	&$22.17$	&$23.85$	&2	&2.1	&0.3	&0\\
SDSS~J1143$+$3017	&$11^\mathrm{h}43^\mathrm{m}14\fs67$	&$+30\degr17\arcmin11\farcs8$	&$2.964$	&$18.89$	&$r$	&$21.66$	&$23.43$	&2	&2.5	&0	&0\\
SDSS~J2039$-$0047	&$20^\mathrm{h}39^\mathrm{m}06\fs09$	&$-00\degr47\arcmin36\farcs6$	&$2.966$	&$19.46$	&$r$	&$23.40$	&$26.03$	&2	&3.2	&0.5	&0\\
SDSS~J1335$+$2230	&$13^\mathrm{h}35^\mathrm{m}03\fs67$	&$+22\degr30\arcmin52\farcs7$	&$2.972$	&$18.95$	&$r$	&$21.42$	&$21.43$	&0	&2.7	&3.2	&0\\
SDSS~J1356$+$0556	&$13^\mathrm{h}56^\mathrm{m}20\fs83$	&$+05\degr56\arcmin19\farcs7$	&$2.973$	&$19.03$	&$r$	&$21.73$	&$21.77$	&0	&2.6	&2.4	&0\\
2QZ~J0239$-$2749	&$02^\mathrm{h}39^\mathrm{m}23\fs60$	&$-27\degr49\arcmin30\farcs8$	&$2.982$	&$20.11$	&$b_J$	&$23.46$	&$25.07$	&2	&3.3	&0.5	&2\\
SDSS~J2310$+$0048	&$23^\mathrm{h}10^\mathrm{m}55\fs32$	&$+00\degr48\arcmin17\farcs1$	&$2.993$	&$18.71$	&$r$	&$22.66$	&$21.44$	&0	&5.5	&12.2	&0\\
2QZ~J2343$-$2947	&$23^\mathrm{h}43^\mathrm{m}35\fs21$	&$-29\degr47\arcmin00\farcs6$	&$2.995$	&$19.65$	&$b_J$	&$22.27$	&$22.08$	&0	&2.4	&4.4	&2\\
SDSS~J1311$+$0857	&$13^\mathrm{h}11^\mathrm{m}27\fs42$	&$+08\degr57\arcmin15\farcs0$	&$3.009$	&$19.19$	&$r$	&$21.64$	&$23.82$	&2	&2.3	&0	&0\\
SDSS~J1040$+$2446	&$10^\mathrm{h}40^\mathrm{m}03\fs62$	&$+24\degr46\arcmin53\farcs0$	&$3.012$	&$19.46$	&$r$	&$22.25$	&$22.11$	&0	&2.3	&2.1	&0\\
SDSS~J0858$+$4012	&$08^\mathrm{h}58^\mathrm{m}33\fs02$	&$+40\degr12\arcmin03\farcs1$	&$3.013$	&$18.81$	&$r$	&$22.35$	&$22.87$	&1	&2.4	&2.4	&0\\
SDSS~J1146$+$2306	&$11^\mathrm{h}46^\mathrm{m}09\fs81$	&$+23\degr06\arcmin13\farcs7$	&$3.013$	&$18.94$	&$r$	&$21.44$	&$21.13$	&0	&2.6	&3.7	&0\\
SDSS~J2334$-$1039	&$23^\mathrm{h}34^\mathrm{m}49\fs48$	&$-10\degr39\arcmin41\farcs0$	&$3.019$	&$19.96$	&$r$	&$23.14$	&$23.76$	&0	&4.0	&2.2	&1\\
SDSS~J0924$+$4852	&$09^\mathrm{h}24^\mathrm{m}47\fs35$	&$+48\degr52\arcmin42\farcs8$	&$3.020$	&$18.31$	&$r$	&$21.60$	&$21.19$	&0	&2.8	&4.1	&0\\
SDSS~J0947$+$1421	&$09^\mathrm{h}47^\mathrm{m}34\fs19$	&$+14\degr21\arcmin16\farcs9$	&$3.030$	&$17.22$	&$r$	&$20.94$	&$19.70$	&0	&4.2	&9.3	&0\\
SDSS~J1630$+$4145	&$16^\mathrm{h}30^\mathrm{m}05\fs72$	&$+41\degr45\arcmin09\farcs1$	&$3.033$	&$19.53$	&$r$	&$21.20$	&$22.71$	&2	&3.0	&2.2	&0\\
SDSS~J1159$+$3134	&$11^\mathrm{h}59^\mathrm{m}11\fs52$	&$+31\degr34\arcmin27\farcs3$	&$3.055$	&$17.70$	&$r$	&$21.91$	&$21.35$	&0	&2.4	&3.6	&0\\
FIRST~J0921$+$3051	&$09^\mathrm{h}21^\mathrm{m}56\fs27$	&$+30\degr51\arcmin57\farcs1$	&$3.062$	&$18.75$	&$r$	&$21.63$	&$23.94$	&2	&2.8	&0.2	&0\\
SDSS~J1430$+$2307	&$14^\mathrm{h}30^\mathrm{m}06\fs11$	&$+23\degr07\arcmin21\farcs4$	&$3.062$	&$20.16$	&$r$	&$21.54$	&$22.46$	&0	&2.4	&2.9	&0\\
SDSS~J1225$+$1933	&$12^\mathrm{h}25^\mathrm{m}45\fs89$	&$+19\degr33\arcmin41\farcs3$	&$3.066$	&$19.30$	&$r$	&$21.45$	&$21.54$	&0	&3.0	&3.3	&0\\
PMN~J1458$+$0855	&$14^\mathrm{h}58^\mathrm{m}05\fs99$	&$+08\degr55\arcmin30\farcs1$	&$3.066$	&$20.27$	&$r$	&$22.04$	&$23.27$	&2	&2.4	&1.3	&0\\
SDSS~J1207$+$3509	&$12^\mathrm{h}07^\mathrm{m}06\fs99$	&$+35\degr09\arcmin22\farcs2$	&$3.094$	&$19.77$	&$r$	&$21.78$	&$22.19$	&0	&2.3	&3.8	&1\\
SDSS~J1644$+$2143	&$16^\mathrm{h}44^\mathrm{m}39\fs86$	&$+21\degr43\arcmin11\farcs5$	&$3.111$	&$18.43$	&$r$	&$21.60$	&$21.90$	&1	&2.5	&2.9	&1\\
SDSS~J1259$+$6355	&$12^\mathrm{h}59^\mathrm{m}48\fs78$	&$+63\degr55\arcmin36\farcs9$	&$3.114$	&$19.32$	&$r$	&$21.90$	&$22.65$	&2	&2.2	&1.9	&0\\
SDSS~J1215$+$3138	&$12^\mathrm{h}15^\mathrm{m}57\fs28$	&$+31\degr38\arcmin41\farcs4$	&$3.120$	&$20.03$	&$r$	&$21.99$	&$22.37$	&0	&2.4	&2.1	&0\\
SDSS~J1103$+$3629	&$11^\mathrm{h}03^\mathrm{m}25\fs53$	&$+36\degr29\arcmin14\farcs4$	&$3.122$	&$20.36$	&$r$	&$21.99$	&$23.54$	&2	&2.2	&0.5	&0\\
SDSS~J1647$+$2305	&$16^\mathrm{h}47^\mathrm{m}54\fs58$	&$+23\degr05\arcmin15\farcs3$	&$3.136$	&$20.10$	&$r$	&$22.22$	&$24.24$	&2	&2.2	&0	&1\\
SDSS~J0838$+$1924	&$08^\mathrm{h}38^\mathrm{m}33\fs97$	&$+19\degr24\arcmin26\farcs2$	&$3.142$	&$19.43$	&$r$	&$22.97$	&\nodata	&2	&6.6	&\nodata&0\\
FIRST~J1237$+$0126	&$12^\mathrm{h}37^\mathrm{m}48\fs99$	&$+01\degr26\arcmin06\farcs9$	&$3.145$	&$18.88$	&$r$	&$21.66$	&$21.95$	&0	&2.8	&2.1	&0\\
SDSS~J0847$+$1322	&$08^\mathrm{h}47^\mathrm{m}56\fs09$	&$+13\degr22\arcmin02\farcs0$	&$3.147$	&$18.69$	&$r$	&$21.07$	&$22.94$	&2	&3.2	&0.5	&0\\
SDSS~J1416$+$0644	&$14^\mathrm{h}16^\mathrm{m}08\fs43$	&$+06\degr44\arcmin31\farcs8$	&$3.148$	&$18.99$	&$r$	&$22.84$	&$22.42$	&0	&2.6	&3.5	&0\\
SDSS~J0814$+$4846	&$08^\mathrm{h}14^\mathrm{m}09\fs76$	&$+48\degr46\arcmin45\farcs1$	&$3.159$	&$21.28$	&$r$	&$23.80$	&$24.24$	&0	&3.7	&2.3	&0\\
Q~0044$-$273		&$00^\mathrm{h}47^\mathrm{m}10\fs84$	&$-27\degr04\arcmin41\farcs0$	&$3.160$	&$20.20$	&$R$	&$21.68$	&$23.42$	&2	&2.6	&0.4	&2\\
SDSS~J1508$+$1654	&$15^\mathrm{h}08^\mathrm{m}28\fs78$	&$+16\degr54\arcmin33\farcs1$	&$3.172$	&$18.35$	&$r$	&$21.18$	&$22.79$	&2	&3.0	&0.2	&0\\
SDSS~J1251$+$4120	&$12^\mathrm{h}51^\mathrm{m}25\fs36$	&$+41\degr20\arcmin00\farcs4$	&$3.173$	&$18.95$	&$r$	&$24.56$	&$25.65$	&2	&3.2	&1.3	&0\\
SDSS~J1404$+$1248	&$14^\mathrm{h}04^\mathrm{m}04\fs23$	&$+12\degr48\arcmin59\farcs1$	&$3.187$	&$19.23$	&$r$	&$21.99$	&$23.76$	&2	&2.1	&0	&0\\
SDSS~J2345$+$0108	&$23^\mathrm{h}45^\mathrm{m}41\fs56$	&$+01\degr08\arcmin18\farcs2$	&$3.190$	&$19.75$	&$r$	&$22.84$	&$24.32$	&2	&2.4	&0.5	&1\\
QSO~J0332$-$2747	&$03^\mathrm{h}32^\mathrm{m}42\fs84$	&$-27\degr47\arcmin02\farcs5$	&$3.193$	&$24.10$	&$R$	&$25.82$	&$25.17$	&1	&2.4	&3.4	&3\\
SDSS~J0856$+$1234	&$08^\mathrm{h}56^\mathrm{m}33\fs57$	&$+12\degr34\arcmin28\farcs5$	&$3.195$	&$18.68$	&$r$	&$21.85$	&$21.15$	&0	&3.0	&5.4	&0\\
SDSS~J1454$+$3741	&$14^\mathrm{h}54^\mathrm{m}37\fs08$	&$+37\degr41\arcmin34\farcs5$	&$3.195$	&$19.09$	&$r$	&$21.88$	&$23.13$	&0	&2.6	&1.7	&0\\
SDSS~J1000$+$3123	&$10^\mathrm{h}00^\mathrm{m}20\fs25$	&$+31\degr23\arcmin07\farcs0$	&$3.230$	&$20.09$	&$r$	&$21.71$	&$23.06$	&2	&2.2	&0	&0\\
SDSS~J0955$+$4322	&$09^\mathrm{h}55^\mathrm{m}46\fs35$	&$+43\degr22\arcmin44\farcs7$	&$3.240$	&$19.47$	&$r$	&$21.12$	&$21.81$	&0	&3.3	&2.6	&0\\
SDSS~J1352$+$1251	&$13^\mathrm{h}52^\mathrm{m}49\fs76$	&$+12\degr51\arcmin37\farcs0$	&$3.266$	&$18.84$	&$r$	&$21.73$	&$22.32$	&0	&2.1	&1.8	&0\\
SDSS~J1110$+$1804	&$11^\mathrm{h}10^\mathrm{m}07\fs29$	&$+18\degr04\arcmin39\farcs6$	&$3.270$	&$18.36$	&$r$	&$22.31$	&$21.88$	&1	&2.5	&3.8	&0\\
HS~0954$+$3549		&$09^\mathrm{h}57^\mathrm{m}35\fs37$	&$+35\degr35\arcmin20\farcs6$	&$3.277$	&$18.16$	&$r$	&$22.38$	&$21.40$	&0	&2.6	&3.9	&0\\
SDSS~J2313$+$1441	&$23^\mathrm{h}13^\mathrm{m}32\fs22$	&$+14\degr41\arcmin22\farcs4$	&$3.337$	&$19.75$	&$r$	&$23.17$	&$24.23$	&0	&3.2	&2.9	&0\\
SDSS~J0855$+$2932	&$08^\mathrm{h}55^\mathrm{m}03\fs81$	&$+29\degr32\arcmin48\farcs9$	&$3.388$	&$19.10$	&$r$	&$22.09$	&$21.67$	&0	&2.7	&3.2	&1\\
RDS~080A		&$10^\mathrm{h}51^\mathrm{m}44\fs63$	&$+57\degr28\arcmin08\farcs9$	&$3.409$	&$21.20$	&$R$	&$23.44$	&$22.31$	&0	&7.7	&22.7	&3\\
2GZ~J1153$-$0419	&$11^\mathrm{h}53^\mathrm{m}38\fs90$	&$-04\degr19\arcmin53\farcs0$	&$3.410$	&$19.10$	&$b_J$	&$21.53$	&$20.93$	&0	&2.2	&5.9	&2\\
SDSS~J1339$+$0703	&$13^\mathrm{h}39^\mathrm{m}51\fs84$	&$+07\degr03\arcmin05\farcs1$	&$3.438$	&$20.28$	&$r$	&$22.26$	&$22.67$	&2	&2.1	&2.1	&1\\
SDSS~J1334$+$5213	&$13^\mathrm{h}34^\mathrm{m}48\fs70$	&$+52\degr13\arcmin18\farcs0$	&$3.605$	&$18.77$	&$r$	&$21.90$	&$22.48$	&1	&2.4	&2.0	&1\\
Q~1422$+$231		&$14^\mathrm{h}24^\mathrm{m}38\fs09$	&$+22\degr56\arcmin00\farcs5$	&$3.620$	&$15.48$	&$r$	&$21.85$	&$21.81$	&0	&2.0	&4.5	&1\\
SDSS~J1423$+$1303	&$14^\mathrm{h}23^\mathrm{m}25\fs92$	&$+13\degr03\arcmin00\farcs6$	&$5.037$	&$21.23$	&$r$	&$21.96$	&$23.91$	&2	&2.4	&0	&1
\enddata
\tablecomments{76 of the 87 $z_\mathrm{em}>2.78$ quasars were previously suggested as candidate \ion{He}{2} quasars by \citet{syphers09b}.}
\tablenotetext{a}{The sources listed in Table~\ref{candlist1} are not repeated here.}
\tablenotetext{b}{SDSS $r$ AB magnitude if filter is $r$, otherwise Vega magnitude in given filter.}
\tablenotetext{c}{GALEX limit flag. 0: formal two-band detection, 1: $1\sigma$ lower limit in $m_\mathrm{FUV}$, 2: $1\sigma$ lower limit in $m_\mathrm{NUV}$}
\tablenotetext{d}{Neighbor flag. 0: no SDSS source within $r<5\arcsec$ of the quasar, 1: sufficiently red SDSS source within $r<5\arcsec$ of the quasar, 2: quasar not imaged in SDSS DR7, 3: potential source confusion (DIS detection)}
\end{deluxetable*}

\end{document}